\documentclass[aps,prx,reprint,groupedaddress,amsmath,amssymb,superscriptaddress]{revtex4-2}
\usepackage{graphicx}
\usepackage{float}
\usepackage{multirow}
\usepackage{natbib,twoopt}
\usepackage[breaklinks=true]{hyperref}
\usepackage{xcolor}
\usepackage{amssymb}
\usepackage{stackengine,graphicx}
\usepackage{amsmath}
\usepackage{hyperref}
\usepackage{color}
\usepackage{stackengine}
\usepackage{subfigure}
\usepackage{verbatim}
\usepackage{bbold}
\usepackage{stmaryrd} 
\usepackage{overpic}
\usepackage{makecell} 

\newcommand{\blue}[1]{{\color{black} #1}}

\newcommand{\eps}{\varepsilon}


\renewcommand{\v}[1]{\if\noexpand#1\relax \bm{#1} \else \mathbf{#1} \fi}
\newcommand{\op}[1]{\hat{#1}}
\newcommand{\transp}{\mathsf{T}}

\begin{document}

\title{Nonreciprocal lensing and backscattering suppression via magneto-optical nonlocality}

\author{Dmitry Vagin}
\affiliation{School of Physics and Engineering, ITMO University, Saint  Petersburg 197101, Russia}

\author{Maxim A. Gorlach}
\email{m.gorlach@metalab.ifmo.ru}
\affiliation{School of Physics and Engineering, ITMO University, Saint   Petersburg 197101, Russia}

\begin{abstract}
We introduce a special kind of  nonreciprocal electromagnetic response which gives rise to backscattering suppression in the bulk, a long-sought feature in topological photonics, as well as nonreciprocal lensing~-- an effect when the same structure focuses light incident from one direction and defocuses light propagating in the opposite way. We predict this response in spin spirals and in specially designed metamaterials, validating the key predictions.
\end{abstract}

%


\maketitle

{\it Introduction.~--~} Artificial media have uncovered novel ways to manipulate the propagation of light via tailored electromagnetic responses~\cite{Veselago1968,Elef,Caloz,Marques,Capolino}. When the period of the structure is much smaller than the wavelength, the composite can typically be viewed as a continuous medium possessing a set of frequency-dependent effective material parameters. In addition to the well-known permittivity and permeability those could include bianisotropy~\cite{Kong1974,Tretyakov} which results in the constitutive relations of the form
\begin{gather}   \v{D}=\op{\eps}\,\v{E}+\left(\op{\chi}+i\op{\kappa}\right)\v{H}\:,\\
\v{B}=\left(\op\chi-i\op\kappa\right)^\transp\v{E}+\op{\mu}\,\v{H}\:, 
\end{gather}
being a long-standing topic of active research~\cite{Asadchy2018,Yang2025}. 

Even broader range of phenomena arises due to the electromagnetic nonlocality (spatial dispersion), when the effective permittivity of the medium depends not only on the frequency of the wave, but also on the wave vector~\cite{Agranovich1984}. This results in rich and diverse physics~\cite{Belov2003,Silva2014,Luo2016,Shastri2022} which is a way more pronounced in metamaterials compared to the conventional media~\cite{Gorlach2016,Yang2025}.

On the other hand, the properties of the medium also depend on the applied static magnetic field giving rise to magneto-optical phenomena and electromagnetic nonreciprocity~\cite{Landau,Caloz2018,Asadchy2020}. The latter ensures that the same medium transmits light differently in forward and backward direction providing a physical mechanism for optical isolation. 


Interestingly, the interplay of nonlocal and nonreciprocal phenomena remains poorly studied. 
While the effects of magnetospatial dispersion are probed in condensed matter, their magnitude is very small~\cite{Portigal1971,Ivchenko1984,Krichevtsov1998,Kotova2018}.  In contrast, the metamaterial platform holds a promise to enhance these phenomena by orders of magnitude opening a door to the unique optical effects.

In this Letter, we reveal a special type of magneto-optical nonlocality which provides several important functionalities. First, it enables nonreciprocal lensing: focuses light propagating from one direction and defocuses light travelling in the opposite way. Second, the radiation pattern of any dipole source embedded in such medium is strongly non-symmetric thereby reducing backscattering on random defects in the entire bulk of the medium. Below, we examine this physics, identifying the desired response in conical spin spirals and in a specially designed microwave metamaterial.

{\it Symmetry analysis.~--~} We describe the metamaterial in terms of the nonlocal permittivity tensor \(\eps_{ij}(\omega, \v{k})\), where the dependencies on $\omega$ and $\v{k}$ capture the effects of frequency and spatial dispersion, respectively~\cite{Landau}. In addition, we assume a bias field \(\v{H}_0(\v{r})\), odd under time reversal and responsible for nonreciprocity~\cite{Caloz2018}. In the limit of weak spatial dispersion, we expand the permittivity as
%
%
\begin{equation} \label{eq:general spat disp}
    \eps_{ij}(\omega, \v{k}) = \eps_{ij}^\text{(e)}(\omega) + \eps_{ij}^\text{(o)}(\omega) + \alpha_{ijl}^\text{(e)}(\omega)\; k_l + \alpha_{ijl}^\text{(o)}(\omega) \; k_l\:,
\end{equation}
where the superscripts (e) and (o) denote even and odd components with respect to the reversal of the bias field \(\v{H}_0(\v{r}) ~\to~-\v{H}_0(\v{r})\).  Assuming that the medium is lossless, we find \(\eps_{ij}(\omega, \v{k}) = \eps_{ji}^*(\omega, \v{k})\)~\cite{Landau}, and thus
\begin{align}
    \eps_{ij}^\text{(e)} &= {\eps_{ji}^\text{(e)}}^*, & \eps_{ij}^\text{(o)} &= {\eps_{ji}^\text{(o)}}^*, \label{eq:lossless constrains 1}\\
    \alpha_{ijl}^\text{(e)} &= {\alpha_{jil}^\text{(e)}}^*, & \alpha_{ijl}^\text{(o)} &= {\alpha_{jil}^\text{(o)}}^*. \label{eq:lossless constrains 2}
\end{align}
On the other hand, the permittivity tensor is also constrained by Onsager-Kasimir relations~\cite{Landau,Caloz2018} \(\eps_{ij}(\omega, \v{k}, \v{H}_0) = \eps_{ji}(\omega, -\v{k}, -\v{H}_0)\), which yields
%
\begin{align}
    \eps_{ij}^\text{(e)} &= \eps_{ji}^\text{(e)}, & \eps_{ij}^\text{(o)} &= -\eps_{ji}^\text{(o)}, \label{eq:kinetic constrains 1}\\
    \alpha_{ijl}^\text{(e)} &= -\alpha_{jil}^\text{(e)}, & \alpha_{ijl}^\text{(o)} &= {\alpha_{jil}^\text{(o)}}. \label{eq:kinetic constrains 2}
\end{align}    
Combining the conditions Eqs.~\eqref{eq:lossless constrains 1}-\eqref{eq:kinetic constrains 2}, we recover that the tensor  \(\eps_{ij}^\text{(e)}\) is real and symmetric, while \(\eps_{ij}^\text{(o)}\) is imaginary and antisymmetric. These tensors describe such well-known effects as anisotropy and magneto-optics [see Tab.~\ref{tab:spatial dispersion}].

In the same spirit, we find out that the tensor \(\alpha_{ijl}^\text{(e)}\) responsible for reciprocal nonlocality is imaginary and antisymmetric under the permutation of the first two indices. Tensor \(\alpha_{ijl}^\text{(o)}\) captures magneto-optical nonlocality, being real and symmetric with respect to the same pair of indices. This response is odd both under spatial inversion $\mathcal P$ and time reversal $\mathcal T$, being even under the combined $\mathcal{PT}$ operation. \blue{These symmetry properties distinguish it from the conventional magneto-optical phenomena modified by the second-order spatial dispersion, as considered e.g. in Ref.~\cite{Silveirinha2015}.}  In Table~\ref{tab:spatial dispersion} we present the number of independent components for each of the above tensors.


To connect $\alpha_{ijl}$ to the well-celebrated bianisotropy effects~\cite{Tretyakov}, we note that  any bianisotropic medium with the permittivity $\op{\eps}$, chirality $\op{\kappa}$ and Tellegen tensor $\op{\chi}$ can equivalently be described by the spatially dispersive permittivity~\cite{Kong1990}. Keeping the terms up to the first order in $\v{k}$, we obtain:
\begin{equation} \label{eq:bianis spat disp}
\begin{aligned}
    \op{\eps}(\omega,\v{k}) = \op{\eps}+ \frac{i}{q}\,\left[\op{\kappa}\v{k}^\times+\v{k}^\times\op{\kappa}^\transp\right] + \frac{1}{q}\,\left[\op{\chi}\v{k}^\times - \v{k}^\times\op{\chi}^\transp\right].
\end{aligned}
\end{equation}
Here \(\v{k}^\times\) is a matrix defined as \((\v{k}^\times)_{ij}~=~- \epsilon_{ijl} k_l\), \(\epsilon_{ijl}\) is the Levi-Civita symbol and \(q=\omega/c\).


In bianisotropic framework, imaginary corrections linear in $\v{k}$ arise due to the chirality $\hat{\kappa}$. Since both $\kappa_{ij}$ and $\alpha_{ijl}^\text{(e)}$ have 9 independent components, all reciprocal first-order spatial dispersion effects are properly captured by the chirality tensor [Table~\ref{tab:spatial dispersion}].

At the same time, real corrections linear in $\v{k}$ are related to the Tellegen tensor $\op{\chi}$. In addition, the component of $\op{\chi}$ proportional to the identity matrix does not manifest itself in the bulk, reduces to the boundary term and captures axion electrodynamics~\cite{Wilczek1987, Nenno2020,Shaposhnikov2023,AsadchyNC2024,Yang2025,Liu2025}. This leaves only 8 terms to capture magneto-optical nonlocality. However, the tensor $\alpha_{ijl}^\text{(o)}$ has 18 independent components and hence 10 of them are beyond bianisotropic framework Eq.~\eqref{eq:bianis spat disp}. Those additional components are precisely the focus of our Letter.



\begin{table}[]
    \centering
        \begin{tabular}{|c|c|l|}
        \hline
            Contribution & \makecell{Number of \\ components} & Physical phenomena \\
        \hline    
            \(\eps_{ij}^\text{(e)}\) & 6 & Anisotropy \\
            \(\eps_{ij}^\text{(o)}\) & 3 & Magneto-optics \\
            \(\alpha_{ijl}^\text{(e)}\) & 9 & Electromagnetic chirality\\
            \(\alpha_{ijl}^\text{(o)}\) & 18 & 
            \( \begin{cases}
                \text{Non-reciprocial bianisotropy (8)}\\
                \text{\textbf{Other phenomena} (10)}
            \end{cases}   \) \\
        \hline
        \end{tabular}
    \caption{Different contributions to the nonlocal permittivity tensor and underlying physical effects.}
    \label{tab:spatial dispersion}
\end{table}

{\it Quasi-moving spatial dispersion.~--~} Below, we focus on the specific type of magneto-optical nonlocality \blue{described by the components $\alpha^\text{(o)}_{xzx}=\alpha^\text{(o)}_{zxx}=\alpha^\text{(o)}_{yzy}=\alpha^\text{(o)}_{zyy}$ and captured} by the tensor
\begin{equation} \label{eq:quasi moving spat disp}
    \hat{\eps}(\omega, \v{k}) = \hat{\eps}(\omega, 0) + \frac{\chi_\text{qm}}{q} \begin{pmatrix}
        0 & 0 & k_x \\
        0 & 0 & k_y \\
        k_x & k_y & 0 
    \end{pmatrix}\:,
\end{equation}
where $\chi_{\text{qm}}$ is the dimensionless coefficient quantifying the strength of the effect. It is instructive to compare the form of the nonlocal contribution Eq.~\eqref{eq:quasi moving spat disp} to another celebrated nonreciprocal phenomenon~-- moving-type bianisotropy, which describes the propagation of light in a liquid moving with velocity ${\bf v}=\left(0,0,v_z\right)^T$ relative to the laboratory frame. In the latter case, nonreciprocity tensor takes the form~\cite{Kong1990}
%
\[
    \op{\chi} = \chi_\text{m} \begin{pmatrix}
        0  & 1 & 0 \\
        -1 & 0 & 0 \\
        0  & 0 & 0 
    \end{pmatrix}, \quad \chi_\text{m} = - \frac{v_z}{c} (\eps - 1)\:,
\]
which can be recast as the nonlocal correction to the permittivity tensor
%
\begin{equation}
    \hat{\eps}(\omega, \v{k}) = \hat{\eps}(\omega, 0) + \frac{\chi_\text{m}}{q} \begin{pmatrix}
        -2 k_z & 0 & k_x \\
        0 & -2 k_z & k_y \\
        k_x & k_y & 0 
    \end{pmatrix}.
\end{equation}

This suggests some parallels between the physics of the moving media and our nonlocal magneto-optical response which we further term ``quasi-moving".


A characteristic property of the moving medium is the Fresnel drag~--- the difference between the refractive indices of the wave propagating in forward and backward directions. Such difference is maximized when the wave vector is aligned with the velocity ${\bf v}$ of the flow, and decreases to zero when ${\bf k}\bot{\bf v}$. This physics is readily visualized in terms of the isofrequency contours displaying the length of the wave vector $k$ for various directions at a fixed frequency: the contours become shifted relative to the origin.

\begin{figure*}[ht]
	\centering
	\includegraphics[width=0.8\linewidth]{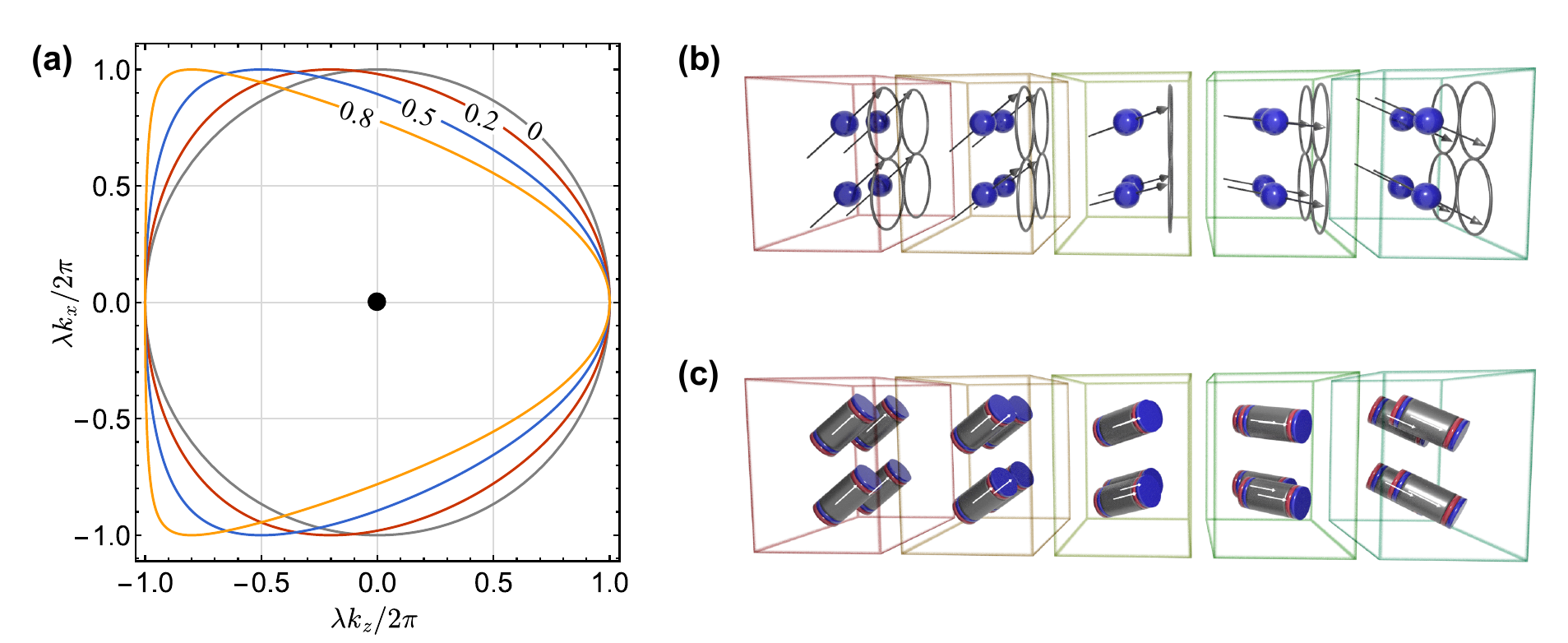}
	\caption{(a) Isofrequency contours for the medium with quasi-moving spatial dispersion. Inversion symmetry of the contours is broken. Different colors encode the values of \(\chi_\text{qm}\) labeled at the curves, \(\lambda\) denotes the wavelength in vacuum. For simplicity, we assume permittivity \(\eps=1\). (b) Spiral arrangement of spins in the lattice enabling quasi-moving spatial dispersion \blue{in such materials as BiFeO$_3$}. (c) \blue{Conceptual metamaterial realization} producing enhanced quasi-moving response. Arrows indicate the direction of the cylinders' magnetization. }
	\label{fig:isofreq}
\end{figure*}

To probe related effects in our system, we solve the dispersion equation (see Supplementary Materials~\cite{Supplement}, Sec.~I) and plot the family of isofrequency contours for the different strengths of $\chi_{\text{qm}}$ [Fig.~\ref{fig:isofreq}(a)]. \blue{While for zero $\chi_{\text{qm}}$ the isofrequency contour is perfectly symmetric,} the contours for $\chi_{\text{qm}}\not=0$ break inversion symmetry in a peculiar manner. If the angle $\theta$ between the wave vector and $Oz$ axis is equal to $0$ or $\pi$, the nonreciprocal effect is not manifested, and forward and backward-propagating waves have the same refractive index. However, the difference arises at oblique incidence enabling an effective angle-dependent Fresnel drag.



{\it Realization.~--~} Importantly, the quasi-moving spatial dispersion occurs both in the natural media and in the engineered metamaterials without the need to modulate their parameters in time. This provides an advantage compared to the moving-type response, which was previously shown to occur in specially designed metasurfaces~\cite{Radi2016, Asadchy2018, Kodama2024} or in the time-varying materials~\cite{Wang2013, Huidobro2019, Galiffi2022}, which struggle to implement this physics in the visible range.

In contrast, the quasi-moving spatial dispersion is present even in the natural structures, where the combination of exchange and Dzyaloshinskii-Moriya interaction~\cite{Cherkasskii2024} enables helical arrangement of spins known as spin spiral [Fig.~\ref{fig:isofreq}(b)]. We derive a simplified model capturing an effective response of such materials studying an idealized multilayered structure consisting of gyrotropic layers~\cite{Supplement} (Sec.~II), where the material equation for each layer reads
\begin{equation} \label{eq:constit rel local}
    \v{D} = \epsilon \v{E} - i [\v{g} \times \v{E}],
\end{equation}
and the component of gyrotropy vector \(\v{g}\) tangential to the layer plane rotates from layer to layer. Assuming fixed angle \(\beta\) between the gyrotropy vector and the layer normal, we find
\begin{equation} \label{eq:chi from gyrotropy}
    \chi_{\text{qm}} =  \frac{q^3 g^3}{2 b^3 \epsilon } \sin^2 \beta \, \cos \beta,
\end{equation}
where the rotation rate of gyrotropy vector is assumed constant, and  \(b=2\pi/a\) is reciprocal lattice constant. A similar physics occurs if the gyrotropy of the permeability is considered instead:
\begin{equation} \label{eq:constit rel local BH}
    \v{B} = \mu \v{H} - i [\v{g} \times \v{H}].
\end{equation}
The quasi-moving effect in such case reads
\begin{equation} 
    \chi_{\text{qm}} = - \frac{q^3 g^3}{2 b^3 \mu } \sin^2 \beta \, \cos \beta.
\end{equation}

We anticipate that the quasi-moving spatial dispersion naturally occurs in chiral magnets~\cite{Togawa2016}, including conical spin-spirals~\cite{Wuhrer2023}.
Qualitatively, these structures may be approximately described by the simplified model of gyrotropic layers when the wavelength exceeds the lattice constant multiple times. A promising candidate for such physics in the optical range is bismuth ferrite (\(\mathrm{BiFeO_3}\)). This compound has spiral magnetic ordering~\cite{Sosnowska1982}, is transparent for visible light~\cite{Kumar2008} and possesses significant anisotropy~\cite{Rivera1997}.


On the other hand, the strength of the quasi-moving effect can be strongly amplified at microwave frequencies \blue{by harnessing resonant microwave metamaterials made of magnetized yttrium iron garnet (YIG) cylinders and metallic inclusions [see Fig.~\ref{fig:isofreq}(c)]. Previously, these structures proved to be instrumental in probing photonic topological phenomena~\cite{Zhou2020} and realizing exotic versions of nonreciprocal bianisotropic response~\cite{Yang2025,Yang2026}.} In such case, the gyrotropy of YIG cylinders can reach the values up to \(g\approx 1.5\) at resonance, \blue{which provides an estimate $\chi_{\text{qm}}\approx 0.2$.} \blue{Introducing additionally resonant metallic inclusions, one can further enhance the quasi-moving effect up to \(\chi_\text{qm} \approx 0.8\)~\cite{Supplement} (Sec.~III).} Our estimates thus suggest that the quasi-moving spatial dispersion is achievable and sizable in realistic metamaterials. We now proceed to studying optical phenomena it unlocks.


\begin{figure}[ht]
	\centering
	\includegraphics[width=0.9\linewidth]{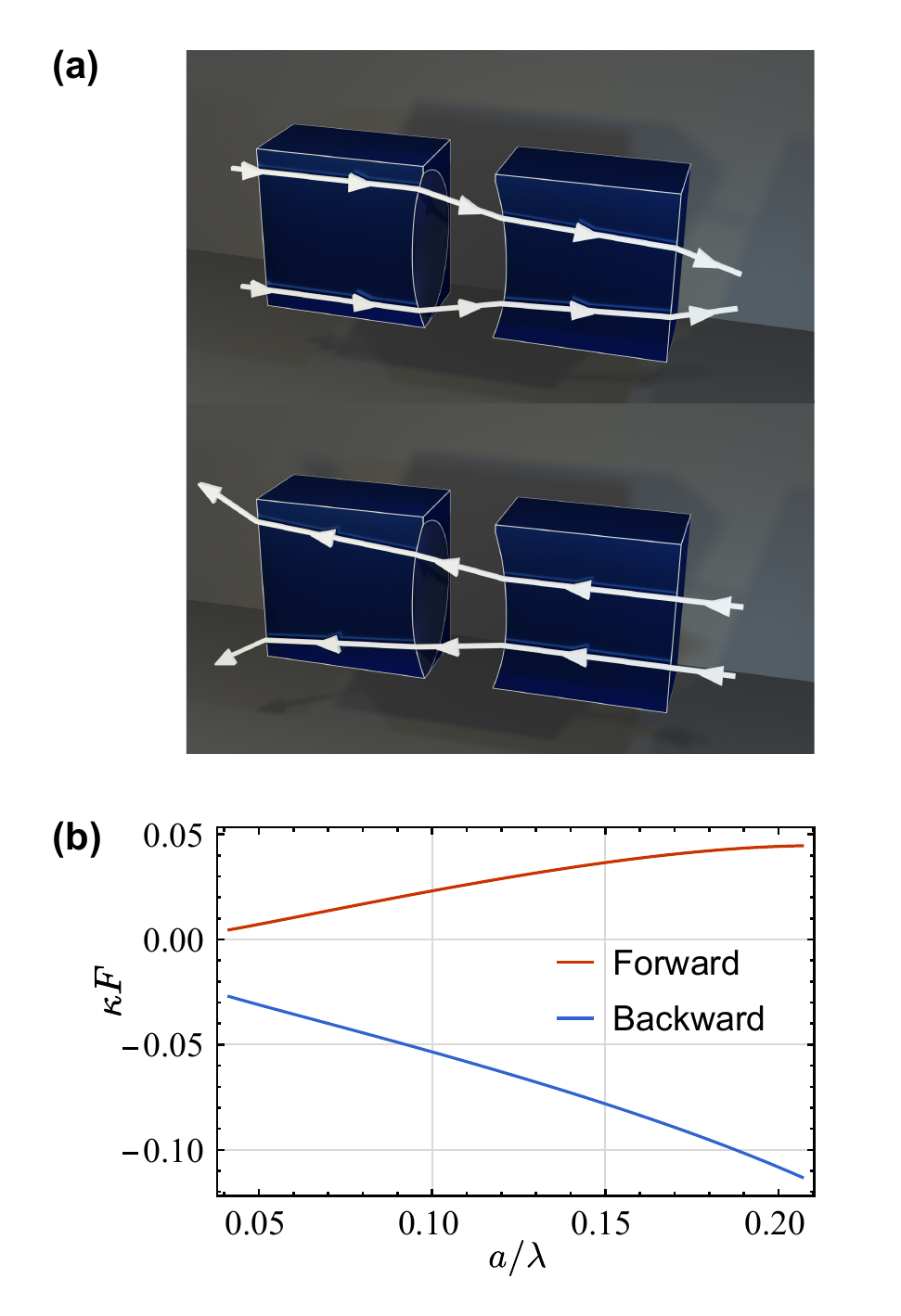}
	\caption{(a) Artistic image of a non-reciprocal lens, with light paths traced for left-to-right and right-to-left propagation.  (b) Focal distance for the fixed light polarization computed for two opposite propagation directions. Material characteristics \(\eps_\perp=2.9\), \(\eps_\parallel=1.0\), \(N=30\), \(g=0.9\), \(\beta=0.3 \pi\). The focal distance is multiplied by the combined curvature $\kappa = 1/R_{1} - 1/R_{2}$ of the refractive surfaces.}
	\label{fig:lens}
\end{figure}

{\it Non-reciprocal lens.~--~} Nonreciprocal nature of the material opens a possibility to construct a lens that has different focal distances for the two opposite directions of light propagation along the $z$ axis. For instance, this system could focus light incident from the left and defocus rays coming from the right [Fig. \ref{fig:lens}(a)]. 

To evaluate the focal distance of a thin lens, we employ the paraxial approximation \(|k_x/k_z| \ll 1 \) and terminate the expansion of the even function \(n(k_x)\) derived from the dispersion equation on the quadratic contribution
\begin{equation}
    n(k_x) = n_\parallel + \zeta \frac{k_x^2}{k^2} = n_\parallel + \zeta \sin^2 \theta.
\end{equation}
Here \(\theta\) is the angle between the wave vector and the axis $Oz$,
\begin{equation}
    \zeta = \mp \chi_\text{qm} - \frac{\chi_\text{qm} ^2}{2 \sqrt{\eps}}\:,
\end{equation}
\blue{where upper and lower signs correspond to the waves propagating forward ($k_z>0$) and backward ($k_z<0$), respectively, see Sec.~IV~\cite{Supplement}.} The dependence of effective refractive index on the angle $\theta$ modifies the optical paths for the rays traversing the system and hence affects the location of the focal spot. Since $\chi_{\text{qm}}$ is inversion-odd, the $\zeta$ term has different values for the light propagating in forward and backward directions. This opens a prospect of non-reciprocal lensing.


Expanding the expression for the optical path length and keeping the terms up to $\theta^2$, we recover the standard thin lens formula~\cite{Supplement} (Sec.~IV) \(1/F = 1/{d_\text{o}} + 1/{d_\text{i}}\), where \(d_\text{o}\) and \(d_\text{i}\) denote distances from the lens to an object and its image, respectively. However, the expression for the optical power is modified:
\[
    \frac{1}{F} = \frac{n_{2 \parallel} - n_{1 \parallel}}{n_{1 \parallel} + 2 \zeta_1} \left(\frac{1}{R_{1}} - \frac{1}{R_{2}}\right).
\]
Here \(R_{1}\) and \(R_{2}\) are curvature radii of the refractive surfaces, \(n_{1 \parallel}\) and \(\zeta_1\) are the properties of the medium surrounding the lens, while \(n_{2 \parallel}\) corresponds to the lens itself. The result does not depend on \(\zeta_2\). Hence, a promising configuration is a lens-shaped cavity cut into two slabs of the quasi-moving material, Fig.~\ref{fig:lens}(a).


By the proper choice of parameters, we ensure that this lens can focus and defocus light depending on the direction it propagates. In Figure~\ref{fig:lens}(b) we plot the focal distances, corresponding to the fixed light polarization: right circular polarization in the case of normal incidence and its continuous modification for oblique incidence. 
The focal distances for the two illumination directions are profoundly different and even have different signs.
A similar physics occurs for another polarization with weaker nonreciprocal effect and the same sign of the focal distances for both propagation directions. An additional property of this system is particularly small focal length per given curvature. This contrasts with the performance of the lenses made from isotropic materials with positive refractive index, where the quantity \(\kappa F\) is always larger than 1.

\begin{figure}[t]
	\centering
	\includegraphics[width=0.7\linewidth]{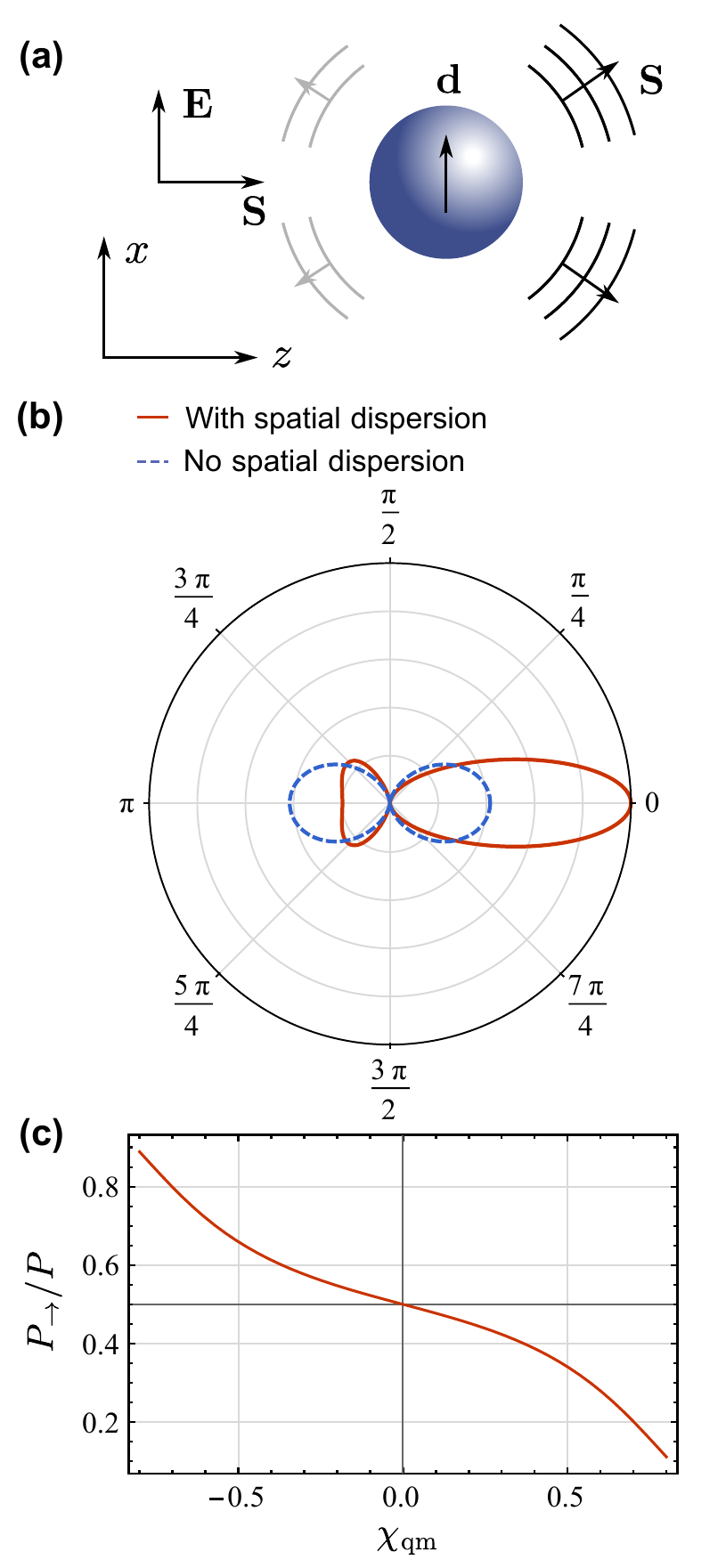}
	\caption{Asymmetric scattering in the quasi-moving medium: (a) A schematic image of a random defect excited by an electromagnetic wave and radiating mostly forward.  (b) Radiation pattern for the \(x\)-aligned dipole in quasi-moving medium with \(\chi_\text{qm}=-0.2\) shown in $Oxz$ plane. (c) The ratio of power radiated in the forward \(+z\) direction to the total radiation power versus the strength of  the quasi-moving effect \(\chi_\text{qm}\).}
	\label{fig:scattering}
\end{figure}

{\it Backscattering suppression.~--~} Another important application of the quasi-moving effect is the control of the radiation power distribution. The far-field analysis based on the Fourier transformation and stationary phase approximation demonstrates~\cite{Supplement} (Sec.~V) that an oscillating dipole radiates asymmetrically in the quasi-moving environment, Fig.~\ref{fig:scattering}(a). The typical radiation pattern of a dipole oscillating in the direction orthogonal to the spin spiral axis calculated for the case of microwave metamaterial with estimated $\chi_{\text{qm}}=-0.2$ is presented in Fig.~\ref{fig:scattering}(b). To quantify the asymmetry of the radiation pattern, we calculate the ratio between the power radiated to the semi-space $z>0$ $P_\rightarrow$ and the total scattered power $P$ in Fig.~\ref{fig:scattering}(c). In particular, experimentally achievable value \(\chi_\text{qm} \approx -0.2\) leads to the ratio around 0.55. \blue{Higher $\chi_{\text{qm}}=-0.8$ results in a stronger asymmetry ratio close to 0.9.} Asymmetry in the radiation pattern persists for other dipole orientations as well~\cite{Supplement} (Sec.~V).

Hence, if the wave propagating in the quasi-moving material in $+z$ direction encounters a defect, only a minor part of the wave energy will be backscattered. Notably, such backscattering immunity is guaranteed in the entire bulk of the material. \blue{This contrasts with the behavior of traditional topological materials, where backscattering immunity occurs only in the region occupied by the topological edge modes. The proposed mechanism is also qualitatively different from the recently developed topological body waveguides~\cite{Chen2022,Chan2026} which operate at frequencies of partial photonic bandgap and require structuring in the transverse direction.}


In summary, we have shown that the interplay of nonlocality and magneto-optical nonreciprocity enables rich optical phenomena which include nonreciprocal lensing and backscattering suppression in the bulk of the medium. This physics is beyond the bianisotropic framework and until now remained vastly unexplored. At the same time, \blue{a practical metamaterial design presented in the End Matter demonstrates that strong quasi-moving response is well within the reach of modern experiments and can readily be realized experimentally in the microwave metamaterials}. This promises fruitful applications in nonreciprocal routing of light and topological photonics.

\begin{acknowledgments}
Theoretical models were supported by the Russian Science Foundation, grant No.~26-72-10054. Numerical simulations were supported by Priority 2030 Federal Academic Leadership Program.


\end{acknowledgments}

\bibliography{NonReciprocalLens-rev1}

\newpage

\section*{End matter}

\blue{
In this section, we put forward a practical metamaterial enabling strong quasi-moving response at microwave frequencies. Performing full-wave numerical simulations, we demonstrate strong asymmetry of transmission in forward and backward directions and the effect of backscattering suppression on defects.

\begin{figure}[b]
	\centering
	\includegraphics[width=0.75\linewidth]{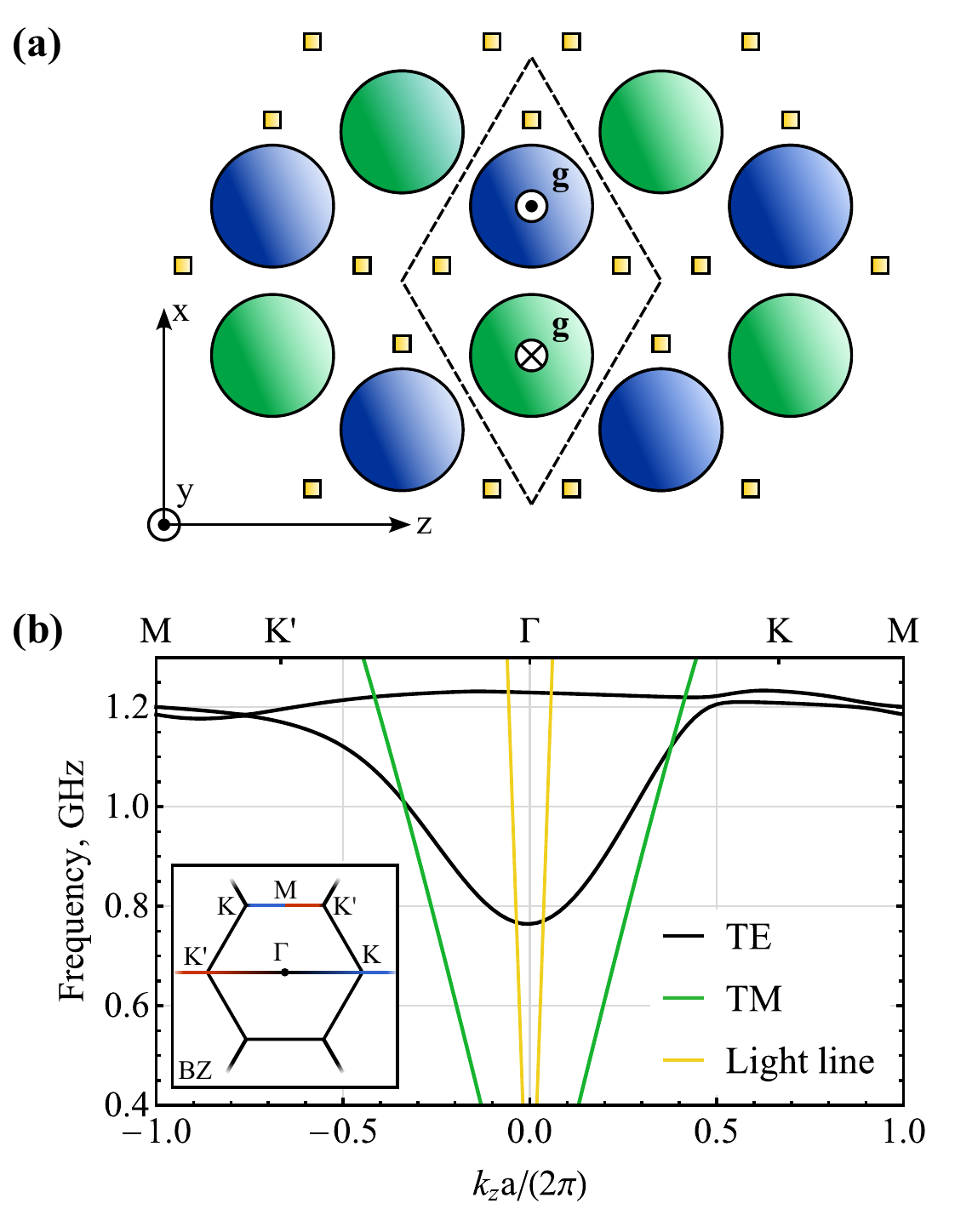}
	\caption{\blue{(a) A practical metamaterial design realizing quasi-moving response and backscattering suppression at microwave frequencies. YIG cylinders with the opposite magnetizations are depicted by blue and green, respectively. Metallic wires are shown in yellow. The dashed rhombus depicts the boundaries of a unit cell. (b) Simulated dispersion diagram of the metamaterial. Nonreciprocity is manifested in the different slopes of TE dispersion curves for positive and negative $k_z$ and in the different bandgaps for forward and backward propagation. Inset depicts the first Brillouin zone.}}
	\label{fig:design}
\end{figure}

Designing a metamaterial, we build on the three considerations: (i) the magnetization distribution does not have to form a spiral and should be ideally collinear to simplify the implementation. Essential symmetry requirements include $\mathcal{P}$ breaking and $\mathcal{T}$ breaking; (ii) at microwave frequencies it is preferable to exploit gyrotropic permeability of yttrium iron garnet (YIG) instead of gyrotropic permittivity. This provides magnetic counterpart of the system discussed in the main text and enables the same optical effects; (iii) we aim to achieve backscattering suppression only for a single light polarization. Achieving the same functionality for both polarizations is feasible, but requires more complicated structure.

The designed metamaterial is shown in Fig.~\ref{fig:design}(a).  It consists of oppositely magnetized YIG rods and electric perfectly conducting wires introduced to resonantly amplify non-reciprocal effects. Lattice constant of the material is $a=11.4$~cm, the radius of each cylinder is $1.55$ cm, wire thickness is $0.2$~cm,  wires in a unit cell form an isosceles triangle with the base equal to $5.7$~cm and height $3.8$~cm.  This system breaks reflection symmetry in $Oyz$ plane and time-reversal symmetry $\mathcal{T}$ due to the magnetization of YIG rods, while the reflection in $Oxz$ is preserved. For simplicity, we assume translational invariance along $y$ axis.


\begin{figure}[b]
	\centering
	\includegraphics[width=0.72\linewidth]{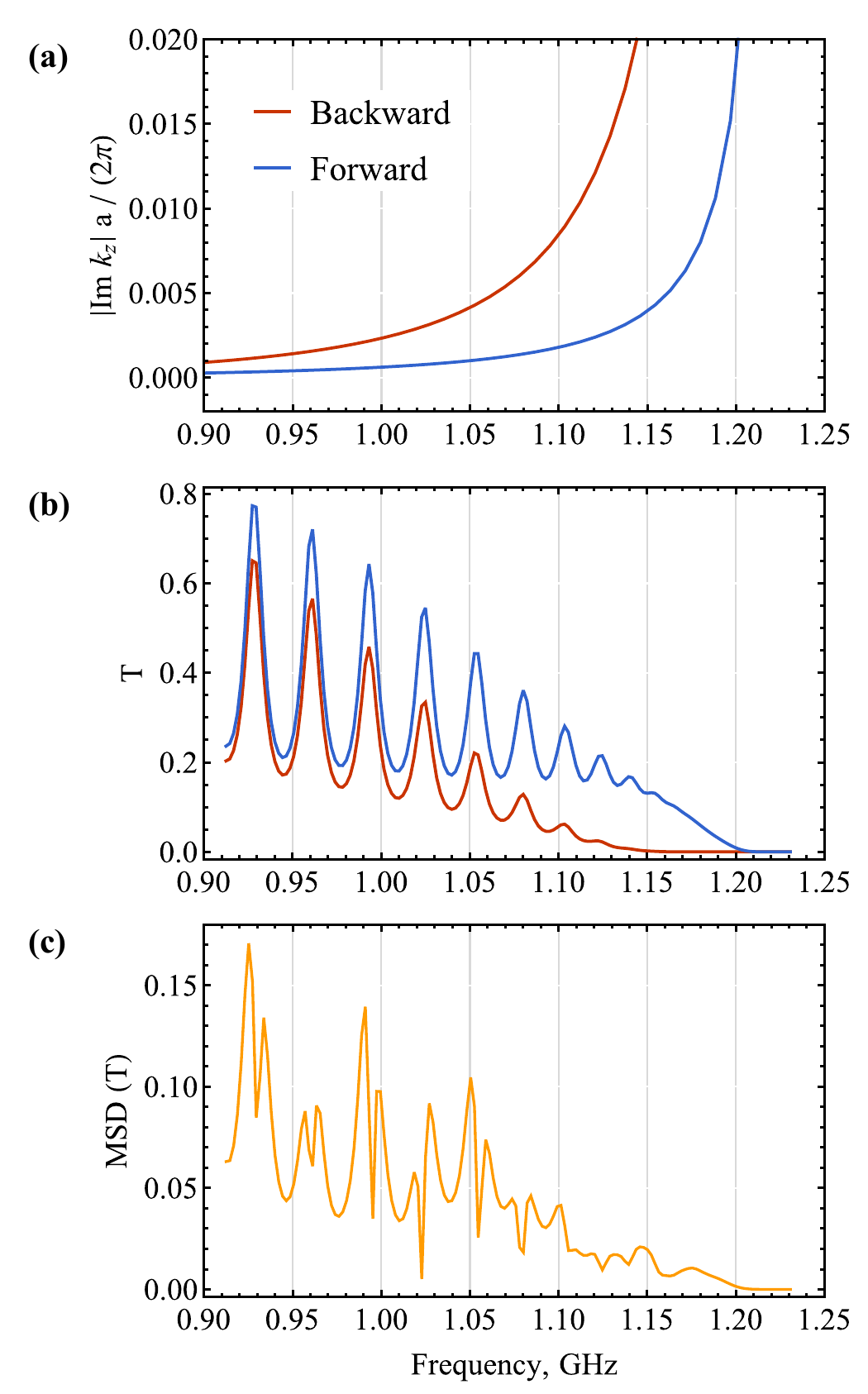}
	\caption{\blue{(a) Imaginary parts of the propagation constant for the waves traveling in forward and backward directions. The forward-propagating wave is attenuated much less than the backward-propagating one. (b) Transmittance $T=|t|^2$, where $t$ is the amplitude transmission coefficient, for a finite sample containing $N=16$ unit cells in forward and backward directions (blue and red lines, respectively). Due to nonreciprocal losses, the system transmits light differently in forward and backward directions. (c) Root mean squared deviation from the values of forward transmittance caused by the defects. At frequencies where backward transmittance is suppressed the influence of the defects becomes negligible.}}
	\label{fig:backscatt suppr}
\end{figure}

To model this structure, we approximate the gyrotropic frequency-dependent YIG permeability by the expressions~\cite{Yang2025}
\begin{align*}
    \op{\mu} &= \begin{pmatrix}
        \mu_r & -i g & 0 \\
        i g & \mu_r & 0 \\ 
        0 & 0 & 1
    \end{pmatrix},     \\
    \mu_r &= 1 + \frac{(\omega_0 + i \alpha \omega) \omega_m}{(\omega_0 + i \alpha \omega)^2 - \omega^2}, \\
    g &= \frac{\omega \omega_m}{(\omega_0 + i \alpha \omega)^2 - \omega^2},
\end{align*}
where the parameters of yttrium iron garnet (YIG) read: $\alpha=0.0088$, $\omega_0 = 2\pi \cdot 1.20 \text{ GHz}$, $\omega_m = 2\pi \cdot 62.6~\text{MHz}$ and permittivity is assumed to be constant: $\eps = 13$. Chosen resonance frequency $\omega_0$ is reached with an external magnetic field of 0.22~T, attainable with commercially available magnets.

Qualitatively, the quasi-moving response for TE-polarized waves  ($E_y\not=0$, $H_y=0$) in this structure can be understood studying a single dimer in a unit cell, Fig.~\ref{fig:design}(a). The incident TE wave has $H_z$ field component. Due to gyrotropy $g$, it induces magnetization of YIG cylinders in the $x$ direction. For a pair of adjacent cylinders with the opposite gyrotropies $x$-directed magnetizations nearly cancel, and the net magnetization is proportional to the phase delay between $H_z$ fields acting on the cylinders, i.e. $k_x$. We thus arrive to the conclusion that
\begin{equation}
M_x\propto k_x\,g\,H_z\:.
\end{equation}
In the effective medium description and for small angles between the wave vector and $z$ axis, this translates into
\begin{equation}
 \hat{\mu} (\mathbf{k}) = \hat{\mu} (k_z) + \frac{\chi_\text{qm}}{q} \begin{pmatrix} 0 &  0 & k_x \\ 0 & 0 & 0  \\  k_x & 0 & 0\end{pmatrix},    
\end{equation}
which is the magnetic analog of the quasi-moving effect studied in the main text. This qualitative reasoning is further supported by calculating the isofrequency contours of the designed metamaterial (see  Sec.~III of Ref.~\cite{Supplement}), which closely match those expected for the quasi-moving medium. The estimated magnitude of the quasi-moving response reaches $\chi_{\text{qm}}\approx 0.8-0.1i$ at 1.1~GHz, suggesting strong nonreciprocity.


To probe the associated effect of backscattering suppression, we proceed by simulating the dispersion diagram of the metamaterial. Given symmetry plane $Oxz$, all modes are either even (TM) or odd (TE) under reflection in $Oxz$. TM modes with magnetic field along the $y$ axis do not experience any nonreciprocity featuring a trivial spectrum shown by the green lines in Fig.~\ref{fig:design}(b). In contrast, the dispersion of TE modes shown by the black lines in Fig.~\ref{fig:design}(b) is strongly non-symmetric under $k_z\rightarrow-k_z$ reflection. From the symmetry point of view, this is guaranteed by the simultaneous breaking of the time-reversal symmetry and mirror reflection in the $Oxy$ plane. Note that the same mechanism of simultaneous $\mathcal{P}$ and $\mathcal{T}$ breaking ensures nonreciprocal bands in the other physical contexts, e.g. in the arrays of polarized quantum emitters~\cite{Tang2022}.

Nonsymmetric dispersion in Fig.~\ref{fig:design}(b) results in the different propagation in forward and backward directions. Due to losses, the propagation constants $k_z$ acquire an imaginary contribution shown in Fig.~\ref{fig:backscatt suppr}(a) and characterizing wave attenuation with the distance. The effect of losses on forward and backward propagation appears markedly different, and in the range of frequencies from 1.0 to 1.2 GHz the backward-propagating waves decay much faster compared to the forward-propagating ones.


This immediately translates into nonreciprocal transmission depicted in Fig.~\ref{fig:backscatt suppr}(b). Due to the nonreciprocal nature of losses, the transmission in backward direction is significantly suppressed compared to the forward propagation scenario, especially in the range from 1.1 to 1.15~GHz. This provides a mechanism to suppress undesired backscattering. Indeed, if the wave encounters a defect, it could be scattered either in forward, or in backward direction in this planar geometry. However, backward transmission nearly vanishes, which naturally limits backscattering.

To quantify the effect of backscattering suppression, simulate four representative examples of different defects introduced in the structure:
\begin{enumerate}
    \item Increasing dimension of one wire in $Ox$ direction by a factor of 5.
    \item Decreasing a radius of the cylinder by $40\%$.
    \item Shifting a wire by 0.15 lattice constant in $Oz$ direction,
    \item Shifting a cylinder by 0.1 lattice constant in $Oz$ direction.
\end{enumerate}
For each of the above disorders, we simulate the transmission spectra in forward direction again and compute the root mean square deviation from the transmittance $T$ of  an ideal periodic sample:
\[
    \text{MSD} (T) = \sqrt{\frac{1}{n} \sum_{i=1}^n (T_i - T)^2},
\]
where \(T_i\) is forward transmittance of a sample with a defect number \(i\) out of \(n\) realizations. The calculated result presented in Fig. \ref{fig:backscatt suppr}  (c) shows that in the frequency range 1.1-1.15~GHz the transmittance stays almost unaffected by the defects. From the quantitative side, the transmittance is changed by nearly 20\% due to defects at 0.9 GHz and only by 5\% at 1.15 GHz, while the absolute influence of defects on forward transmittance is reduced by the factor of 10, providing an evidence of sizable backscattering suppression. The influence of defects on reflection is suppressed in the same frequency range (Sec.~VI, Ref.~\cite{Supplement}).



}

\end{document}


\title{Supplementary Materials: \\ Nonreciprocal lensing and backscattering suppression via magneto-optical nonlocality}

\author{Dmitry Vagin}
\affiliation{School of Physics and Engineering, ITMO University, Saint  Petersburg 197101, Russia}

\author{Maxim A. Gorlach}
\email{m.gorlach@metalab.ifmo.ru}
\affiliation{School of Physics and Engineering, ITMO University, Saint  Petersburg 197101, Russia}

\maketitle
\widetext
\tableofcontents

\section{Dispersion equation and eigenmodes in a quasi-moving medium}

Refractive indices and polarizations of the eigenmodes in the medium are essential in the analysis of light propagation. Here we examine the case of quasi-moving nonreciprocal medium. We start from the equation for the electric field of the eigenmode~\cite{Agranovich1984}
\[
    \left(k^2 \op{1}-\v{k} \otimes \v{k} - q^2 \eps(\omega, k) \right) \v{E} = 0.
\]
Using the expression for the nonlocal permittivity of the quasi-moving medium and setting \(k_y=0\), which is possible due to the rotational symmetry around the axis \(z\), we obtain
\[
    \left(
\begin{array}{ccc}
 k_z^2-q^2 \eps  & 0 & -q \chi_\text{qm}  k_x-k_z k_x \\
 0 & k_x^2+k_z^2-q^2 \eps  & 0 \\
 -q \chi_\text{qm}  k_x-k_z k_x & 0 & k_x^2-q^2 \eps  \\
\end{array}
\right) \v{E} = 0\:,
\]
where the permittivity $\hat{\eps}(\omega,0)$ is assumed to be isotropic and equal to $\eps$. {The eigenmode structure is largely similar to the modes of a uniaxial crystal: there is an ordinary TE mode, with \(E_x = E_z = 0\), \(E_y \neq 0\), and an extraordinary TM mode polarized in the \((xz)\) plane. The TE mode does not acquire nonreciprocal contribution, its isofrequency contour is given by 
\[
    k_z^{\text{TE}} = \pm \sqrt{q^2 \eps - k_x^2}\:.
\]
The TM mode dispersion is found by equating the determinant of the given matrix to zero, which yields the solutions
\begin{equation} \label{eq:wavevec qm}    
    k_z^{\text{TM}} =  \pm \sqrt{q^2 \eps - k_x^2} \sqrt{ 1 - \frac{\chi_\text{qm} ^2 k_x^2}{q^2 \eps ^2}} - \frac{\chi_\text{qm}  k_x^2}{q \eps}\:.
\end{equation}
We expect birefringence of obliquely incident light on the boundary of the quasi-moving material, since the refractive indices of two eigenmodes differ.
}
Notably, the magnitude of $k_z$ for the opposite propagation directions is different which is due to the nonreciprocal term $\chi_{\text{qm}}$. Interestingly, the asymmetry disappears under the normal incidence when $k_x=0$.

Note that these equations apply to a `pure' quasi-moving effect, without any additinal spatially dispersive terms or anisotropic contributions. Thus, they do not describe a gyrotropic metamaterial in the form of spin spiral, which, aside from quasi-moving spatial dispersion, features magnetooptical and chiral effects. Those lead to the circular polarization of the eigenmodes at normal incidence ($k_x=0$) and near-circular polarization at oblique incidence.

Note also that the magnetic analog of the discussed metamaterial based on gyrotropic permeability instead of gyrotropic permittivity features nontrivial dispersion for the TE waves and trivial spectrum for TM-polarized modes. This instance is analyzed in the End Matter section.

\section{Effective medium treatment of conical spin spirals}

In this section we show how the quasi-moving spatial dispersion arises in a simplified model of a conical spin spiral. While the rigorous analysis requires quantum-mechanical description of interactions between the spins in the lattice, here we utilize a classical model which embodies all underlying symmetries of a solid state structure and provides conceptually clear picture.

In this description, the spins or meta-atoms are replaced by layered structure with the constant magnetization in each layer. The magnetization generates a gyrotropic term in the constitutive relations:
\begin{equation} \label{eq:constit rel local}
    \v{D} = \v{E} - i [\v{g} \times \v{E}]
\end{equation}
Here \(\v{g}\) is parallel to the magnetization, i.e. spins. As we are studying the case of spin spiral, the direction of \(\v{g}\) periodically depends on the coordinate \(z\):
\[
    \v{g} = g \begin{pmatrix}
        \sin \beta \, \sin\frac{2 \pi z}{a} \\ \sin \beta \, \cos\frac{2 \pi z}{a} \\ \cos \beta 
    \end{pmatrix} = \begin{pmatrix}
        g_\perp \, \sin\frac{2 \pi z}{a} \\ g_\perp \, \cos\frac{2 \pi z}{a} \\ g_z
    \end{pmatrix}
\]
where \(\beta\) is the cone angle and \(a\) is the period of spin spiral. As the spins are localized in the sites of the lattice, the actual magnetization is non-uniform. Here, for simplicity, we assume that the vector ${\bf g}$ is uniform in the $Oxy$ plane.

As a result, the problem reduces to investigating the solutions of Maxwell's equations in a planar geometry where the gyrotropy axis rotates with the $z$ coordinate. Due to the translational symmetry of this structure, we can apply Bloch theorem and expand the field as:
\[
    \begin{pmatrix}
        \v{E}(\v{r}) \\ \v{B}(\v{r}) \\ \v{D}(\v{r}) \\ \v{H}(\v{r})
    \end{pmatrix} = \sum_{n=-\infty}^\infty \begin{pmatrix}
        \v{E}_n \\ \v{B}_n \\ \v{D}_n \\ \v{H}_n
    \end{pmatrix} e^{i \v{k}_n \v{r} - \xi i q c t},
\]
where
\[
    \v{k}_n = \xi \v{k} + n \v{b}, \quad \v{b} = \frac{2 \pi}{a} \v{e}_z, \quad q = \frac{\omega}{c}
\]
In our effective medium analysis we assume that wavelength is much larger then the period of the structure which leads to the conditions \(k \ll b\) and \(q \ll b\). Next we introduce a dimensionless parameter \(\xi\), which controls the wavelength scale, and explore the series expansion of the solution near \(\xi=0\). The substitution of Floquet harmonics allows us to recast Maxwell's equation in the form
\begin{equation} \label{eq:maxwell sys}
\eqsys{
    n \v{b} \times \v{H}_n &= \xi (- \v{k} \times \v{H}_n - q \v{D}_n) + \frac{4 \pi}{i c} \v{j} \, \delta_{n, 0} \\
    n \v{b} \cdot \v{D}_n &= \xi( - \v{k} \cdot \v{D}_n) + \frac{4 \pi}{i} \rho \, \delta_{n, 0}\\
    n \v{b} \times \v{E}_n &= \xi (- \v{k} \times \v{E}_n + q \v{B}_n ) \\
    n \v{b} \cdot \v{B}_n &= \xi (- \v{k} \cdot \v{B}_n).
}
\end{equation}
These equations can be rewritten compactly by utilizing a specific grouping of the fields:
\begin{align*}
    \v{F} &= (E_x, E_y, D_z, H_x, H_y, B_z)^\transp \\
    \v{G} &= (D_x, D_y, E_z, B_x, B_y, H_z)^\transp
\end{align*}
By using this substitition we reduce Eq.~\eqref{eq:maxwell sys} to a single matrix equation
\begin{equation} \label{eq:maxwell sys compact}
    \v{F}_n = \frac{\xi}{b n} \left(- k_z \v{F}_n + \op{\eta} \v{G}_n \right), \quad n \neq 0,
\end{equation}
where
\[
    \op{\eta} = \begin{pmatrix}
     0 & 0 & k_1 & 0 & q & 0 \\
     0 & 0 & k_2 & -q & 0 & 0 \\
     -k_1 & -k_2 & 0 & 0 & 0 & 0 \\
     0 & -q & 0 & 0 & 0 & k_1 \\
     q & 0 & 0 & 0 & 0 & k_2 \\
     0 & 0 & 0 & -k_1 & -k_2 & 0 \\
    \end{pmatrix} = \begin{pmatrix}
        [\v{e}_z^\times, \v{k}^\times] & -q \v{e}_z^\times \\ q \v{e}_z^\times & [\v{e}_z^\times, \v{k}^\times]
    \end{pmatrix}.
\]
The series expansion near \(\xi=0\) reads
\begin{equation}
    \v{F}_n = \sum_{\mu=0}^{\infty} \xi^\mu \v{F}^{(\mu)}_n, \quad \v{G}_n = \sum_{\mu=0}^{\infty} \xi^\mu \v{G}^{(\mu)}_n.
\end{equation}
Thus, the relation \ref{eq:maxwell sys compact} can be used to express the next order using the previous one:
\begin{equation} \label{eq:F next order}
    \v{F}^{(\mu + 1)}_n = \frac{1}{b n} \left(- k_z \v{F}^{(\mu)}_n + \op{\eta} \v{G}^{(\mu)}_n \right), \quad n \neq 0.
\end{equation}
Constitutive relations can be expressed via \(\v{F}\) and \(\v{G}\) as well. Due to linearity of the constitutive relations, there exists a matrix \(\op{\iota}\) satisfying \(\v{G} = \op{\iota} \v{F}\). By applying this relation to the Floquet harmonics we find
\begin{equation} \label{eq:constit rel iota}
    \v{G}_n = \sum_{n'} \op{\iota}_{n-n'} \v{F}_{n'},
\end{equation}
where \(\op{\iota}_{n}\) are Fourier coefficients of \(\op{\iota}\),
\[
    \op{\iota}(z) = \sum_n \op{\iota}_n e^{inbz}
\]
Note that the equations for \(n=0\) need not be solved for the derivation of effective response, as it is sufficient to find all fields as functions of \(\v{F}_0\). Constitutive relations and Maxwell's equations are linear, so we expect the linear mapping given by matrices \(\op{F}^{\mu}_n\) and \(\op{G}^{\mu}_n\):
\begin{equation}
    \v{F}^{(\mu)}_n = \op{F}^{(\mu)}_n \v{F}_0, \quad \v{G}^{(\mu)}_n = \op{G}^{(\mu)}_n \v{F}_0.
\end{equation}
The assumption of constant \(\v{F}_0\), together with Maxwell's equations~\eqref{eq:F next order} and constitutive relations~\eqref{eq:constit rel iota} generate a system of recurrence relations:
\begin{equation} \label{eq:next order system}
    \eqsys{
        \op{F}^{(\mu + 1)}_n &= \frac{1}{b n} (- k_z \op{F}^{(\mu)}_n + \op{\eta} \op{G}^{(\mu)}_n ), \quad n \neq 0\\
        \op{F}^{(\mu + 1)}_0 &= 0 \\ 
        \op{G}^{(\mu + 1)}_n &= \sum_{n'} \op{\iota}_{n-n'} \op{F}^{(\mu + 1)}_n
    }
\end{equation}
In the case of material equations with rotating gyrotropy, Eq. \eqref{eq:constit rel local}, the matrix \(\op{\iota}\) reads 
\[
    \op{\iota}(z) = \left(
        \begin{array}{cccccc}
         \eps -\frac{\sin ^2\left(\frac{2 \pi  z}{a}\right) g_\perp^2}{\eps } & \frac{\sin \left(\frac{4 \pi  z}{a}\right) g_\perp^2}{2 \eps }+i g_z & -\frac{i \sin \left(\frac{2 \pi  z}{a}\right) g_\perp}{\eps } & 0 & 0 & 0 \\
         \frac{\sin \left(\frac{4 \pi  z}{a}\right) g_\perp^2}{2 \eps }-i g_z & \eps -\frac{\cos ^2\left(\frac{2 \pi  z}{a}\right) g_\perp^2}{\eps } & \frac{i \cos \left(\frac{2 \pi  z}{a}\right) g_\perp}{\eps } & 0 & 0 & 0 \\
         -\frac{i \sin \left(\frac{2 \pi  z}{a}\right) g_\perp}{\eps } & \frac{i \cos \left(\frac{2 \pi  z}{a}\right) g_\perp}{\eps } & \frac{1}{\eps } & 0 & 0 & 0 \\
         0 & 0 & 0 & 1 & 0 & 0 \\
         0 & 0 & 0 & 0 & 1 & 0 \\
         0 & 0 & 0 & 0 & 0 & 1 \\
        \end{array}
        \right)
\]
Using recurrence relations \eqref{eq:next order system} the following contributions to \(\op{G}_0\) are evaluated (for brevity we omit the terms quadratic in \(\v{k}\)):
\begin{align*}
    \op{G}_0^{(0)} &= \left(
        \begin{array}{cccccc}
         \eps -\frac{g_\perp^2}{2 \eps } & i g_z & 0 & 0 & 0 & 0 \\
         -i g_z & \eps -\frac{g_\perp^2}{2 \eps } & 0 & 0 & 0 & 0 \\
         0 & 0 & \frac{1}{\eps } & 0 & 0 & 0 \\
         0 & 0 & 0 & 1 & 0 & 0 \\
         0 & 0 & 0 & 0 & 1 & 0 \\
         0 & 0 & 0 & 0 & 0 & 1 \\
        \end{array}
        \right), \\
    \op{G}_0^{(1)} &= \left(
        \begin{array}{cccccc}
         0 & 0 & -\frac{i g_\perp^2 k_y}{2 b \eps ^2} & 0 & 0 & 0 \\
         0 & 0 & \frac{i g_\perp^2 k_x}{2 b \eps ^2} & 0 & 0 & 0 \\
         -\frac{i g_\perp^2 k_y}{2 b \eps ^2} & \frac{i g_\perp^2 k_x}{2 b \eps ^2} & 0 & 0 & 0 & 0 \\
         0 & 0 & 0 & 0 & 0 & 0 \\
         0 & 0 & 0 & 0 & 0 & 0 \\
         0 & 0 & 0 & 0 & 0 & 0 \\
        \end{array}
        \right), \\
    \op{G}_0^{(2)} &= \left(
        \begin{array}{cccccc}
         \frac{q^2 g_\perp^4}{16 b^2 \eps ^2} & 0 & 0 & 0 & 0 & 0 \\
         0 & \frac{q^2 g_\perp^4}{16 b^2 \eps ^2} & 0 & 0 & 0 & 0 \\
         0 & 0 & -\frac{q^2 g_\perp^2}{b^2 \eps ^2} & 0 & 0 & 0 \\
         0 & 0 & 0 & 0 & 0 & 0 \\
         0 & 0 & 0 & 0 & 0 & 0 \\
         0 & 0 & 0 & 0 & 0 & 0 \\
        \end{array}
        \right),\\
    \op{G}_0^{(3)} &= \left(
        \begin{array}{cccccc}
         0 & \frac{i q^2 g_\perp^4 k_z}{16 b^3 \eps ^2} & \frac{q^2 g_\perp^2 \left(8 \eps  g_z k_x-i \left(8 \eps ^2-9 g_\perp^2\right) k_y\right)}{16 b^3 \eps ^3} & 0 & 0 & 0 \\
         -\frac{i q^2 g_\perp^4 k_z}{16 b^3 \eps ^2} & 0 & \frac{q^2 g_\perp^2 \left(i \left(8 \eps ^2-9 g_\perp^2\right) k_x+8 \eps  g_z k_y\right)}{16 b^3 \eps ^3} & 0 & 0 & 0 \\
         \frac{i q^2 g_\perp^2 \left(8 i \eps  g_z k_x+\left(9 g_\perp^2-8 \eps ^2\right) k_y\right)}{16 b^3 \eps ^3} & \frac{i q^2 g_\perp^2 \left(\left(8 \eps ^2-9 g_\perp^2\right) k_x+8 i \eps  g_z k_y\right)}{16 b^3 \eps ^3} & 0 & 0 & 0 & 0 \\
         0 & 0 & 0 & 0 & 0 & 0 \\
         0 & 0 & 0 & 0 & 0 & 0 \\
         0 & 0 & 0 & 0 & 0 & 0 \\
        \end{array}
        \right).
\end{align*}
From these expressions we immediately find the effective parameters: since \(\v{G}_0^{(\mu)} = \op{G}_0^{(\mu)} \v{F}_0\), the effective response, connecting 0-th Floquet harmonics, becomes apparent.
\begin{equation}
    \v{G}_0 = \op{\iota}_\text{eff} \v{F}_0 = \sum_{\mu=0}^{\infty} \xi^\mu  \op{G}_0^{(\mu)} \v{F}_0, \quad \op{\iota}_\text{eff} = \sum_{\mu=0}^{\infty} \xi^\mu  \op{G}_0^{(\mu)}
\end{equation}
We observe that \(\op{\iota}\) has the following block form:
\[
    \op{\iota}_\text{eff}= \begin{pmatrix}
        \op{\sigma} & 0 \\
        0 & 1
    \end{pmatrix},
\]
which corresponds to magnetic permeability \(\op{\mu}_\text{eff}=1\) and permittivity
\[
    \op{\eps}_\text{eff} = \begin{pmatrix}
        \sigma _{xx}-\frac{\sigma _{xz} \sigma _{zx}}{\sigma _{zz}} & \sigma _{xy}-\frac{\sigma _{xz} \sigma _{zy}}{\sigma _{zz}} & \frac{\sigma _{xz}}{\sigma _{zz}} \\
         \sigma _{yx}-\frac{\sigma _{yz} \sigma _{zx}}{\sigma _{zz}} & \sigma _{yy}-\frac{\sigma _{yz} \sigma _{zy}}{\sigma _{zz}} & \frac{\sigma _{yz}}{\sigma _{zz}} \\
         -\frac{\sigma _{zx}}{\sigma _{zz}} & -\frac{\sigma _{zy}}{\sigma _{zz}} & \frac{1}{\sigma _{zz}}
    \end{pmatrix}.
\]
After explicit evaluation of \(\op{\eps}_\text{eff}\), we separate it into three components: constant anisotropic and magneto-optic terms, spatially dispersive chiral term and non-reciprocal quasi-moving element:
\begin{equation}
    \op{\eps}_\text{eff} = \op{\eps}_\text{const} + \op{\eps}_\text{chiral} + \op{\eps}_\text{qm}
\end{equation}
Written below are the results for each contribution up to the third order in \(\xi\):
\begin{align}
    \op{\eps}_\text{const} &= \left(
        \begin{array}{ccc}
         \eps -\frac{g_\perp^2}{2 \eps } & i g_z & 0 \\
         -i g_z & \eps -\frac{g_\perp^2}{2 \eps } & 0 \\
         0 & 0 & \eps  \\
        \end{array}
        \right) + \xi^2 \left(
        \begin{array}{ccc}
         \frac{q^2 g_\perp^4}{16 b^2 \eps ^2} & 0 & 0 \\
         0 & \frac{q^2 g_\perp^4}{16 b^2 \eps ^2} & 0 \\
         0 & 0 & \frac{q^2 g_\perp^2}{b^2} \\
        \end{array}
        \right) \\
    \op{\eps}_\text{chiral} &= \xi \left(
        \begin{array}{ccc}
         0 & 0 & -\frac{i g_\perp^2 k_y}{2 b \eps } \\
         0 & 0 & \frac{i g_\perp^2 k_x}{2 b \eps } \\
         \frac{i g_\perp^2 k_y}{2 b \eps } & -\frac{i g_\perp^2 k_x}{2 b \eps } & 0 \\
        \end{array}
        \right) + \xi^3 \frac{i q^2 g_\perp^2 k_y}{16 b^3 \eps ^2}\left(
        \begin{array}{ccc}
         0 & g_\perp^2 k_z & (g_\perp^2 - 8 \eps^2) k_y \\
         -g_\perp^2 k_z & 0 & (g_\perp^2 - 8 \eps^2) k_x \\ 
         -(g_\perp^2 - 8 \eps^2) k_y & -(g_\perp^2 - 8 \eps^2) k_x & 0
        \end{array}
        \right) \\
    \op{\eps}_\text{qm} &= \xi^3 \frac{q^2 g_\perp^2 g_z}{2 b^3 \eps } \begin{pmatrix}
        0 & 0 & k_x \\
        0 & 0 & k_y \\
        k_x & k_y & 0 \\
    \end{pmatrix}
\end{align}
Imaginary Hermitian matrix \(\op{\eps}_\text{chiral}\) can be re-interpreted as electromagnetic chirality. Non-reciprocal term \(\op{\eps}_\text{qm}\) characterizes spatial dispersion which cannot be described with bianisotropic parameters. The corresponding value \(\chi_\text{qm}\) is found to be
\[
    \chi_\text{qm} = \frac{q^3 g_\perp^2 g_z}{2 b^3 \eps } = \frac{q^3 g^3}{2 b^3 \eps } \sin^2 \beta \, \cos \beta.
\] 

A similar result can be obtained when magnetic permeability is gyrotropic, i. e. 
\[
    \v{B} = \v{H} - i [\v{g} \times \v{H}].
\]
In this case the calculation for \(\op{\iota}_\text{eff}\) is carried out in a similar manner, and, again, the matrix \(\op{\iota}\) has a block-diagonal form, acquiring a spatially-dispersive contribution in some components. In this case we have nontrivial magnetic permeability:
\begin{equation}
    \op{\mu} = \op{\mu}_\text{const} + \op{\mu}_\text{chiral} + \op{\mu}_\text{qm},
\end{equation}

\begin{equation}
    \op{\mu}_\text{qm} = - \xi^3 \frac{q^2 g_\perp^2 g_z}{2 b^3 \mu } \begin{pmatrix}
        0 & 0 & k_x \\
        0 & 0 & k_y \\
        k_x & k_y & 0 \\
    \end{pmatrix}. 
\end{equation}
Here the magnitude of spatial dispersion reads
\[
    \chi_\text{qm} = -\frac{q^3 g_\perp^2 g_z}{2 b^3 \mu } = -\frac{q^3 g^3}{2 b^3 \mu } \sin^2 \beta \, \cos \beta.
\]
If the mapping to spatially-dispersive permittivity is applied, we obtain third-order corrections in \(\v{k}\):
\[
    \eps(q, \v{k}) = 1 + \frac{1}{q^2} \v{k}^\times(\mu^{-1} - \op{1})\v{k}^\times
\]
Despite the difference in the effective permittivity, shown by this derivation, physically magnetic and electric quasi-moving dispersion result in the same optical phenomena and the duality transformation provides a map between them.

In the main text, we examine the implications of this nonreciprocal effect, applicable to both electric and magnetic-type quasi-moving dispersion.

\section{Estimates of the quasi-moving effect in microwave metamaterials}

In this section, we estimate the magnitude of the quasi-moving spatial dispersion available in microwave metamaterials. For the estimates, we employ the parameters of Ref.~\cite{Zhou2020} which studied photonic anti-chiral edge states in a microwave structure composed of magnetized YIG cylinders. The permeability of yttrium iron garnet (YIG) cylinders is approximated by 
\[
    \op{\mu} = \begin{pmatrix}
        \mu_r & -ig & 0 \\ ig &\mu_r &0 \\ 0 & 0 & 1\end{pmatrix},
\]
where all parameters depend on the operating frequency \(\omega\) as
\[
    \mu_r = 1 + \frac{(\omega_0 + i \alpha \omega) \omega_m} {(\omega_0 + i \alpha \omega)^2 - \omega^2}, \quad
    g = \frac{\omega \omega_m}{(\omega_0 + i \alpha \omega)^2 - \omega^2}.
\]
The frequences \(\omega_0 = \gamma \mu_0 H_z\), \(\omega_m = \gamma \mu_0 M_s\) in these dependencies are functions of external magnetic flux density \(\mu_0 H_z = 0.043\,\text{T}\) and magnetic flux density from saturation magnetization \(\mu_0 M_s = 0.00224\,\text{T}\). \(\gamma\) denotes gyromagnetic ratio and \(\alpha =0.0088\) is the damping coefficient. This dependency peaks at \(\Re g= 1.5\) near the resonance \(\omega^2 \approx \omega_0^2\). We apply the expression 
\[
    \chi_\text{qm} =  \frac{1}{2} \left(\frac{2 \pi \omega a}{c} \right)^3 g^3 \sin^2 \beta \cos \beta
\]
to a metamaterial with a period \(a=0.12\,\text{m}\). At resonance, we obtain \(2 \pi \omega a/c \approx 0.5\), where metamaterial approximation has some problems, but remains more or less valid. The maximum with respect to the cone angle \(\beta\) is achieved when \(\beta=\arcsin\sqrt{2/3}\), in this case \(\sin^2 \beta \cos \beta = 2/(3\sqrt{3})\). With all this, we arrive to an estimate
\[
    \chi_\text{qm} \approx 0.2.
\]

It should be stressed that this is just an order of magnitude estimate, and the effect can be enhanced further by properly optimizing the geometry.


{To further optimize the design and enhance nonreciprocity, we incorporate electric conductors as a means to concentrate the field. We outline the details of proposed realistic metamaterial in the End Matter section. We extract the strength of the quasi-moving response from numerical simulation by analyzing the shape of isofrequency contours presented in Fig.~\ref{fig:sim isofreq}. The simulated points are compared to the analytical expression \eqref{eq:wavevec qm}. Fitting both functions \(k_x(\Re k_z)\) and \(\Im k_z(\Re k_z)\) via least-squares method numerically, we extract the value of the complex parameter \(\chi_\text{qm}\).

The isofrequency contour after optimization closely matches the simulated data [Fig. \ref{fig:sim isofreq}] for the frequency 1.15 GHz. Some deviations occur because effective approximation is not fully applicable at this frequency (\(a/\lambda \approx 0.46\)), and higher-order effects arise. 

\begin{figure}
    \centering
    \includegraphics[width=0.4\linewidth]{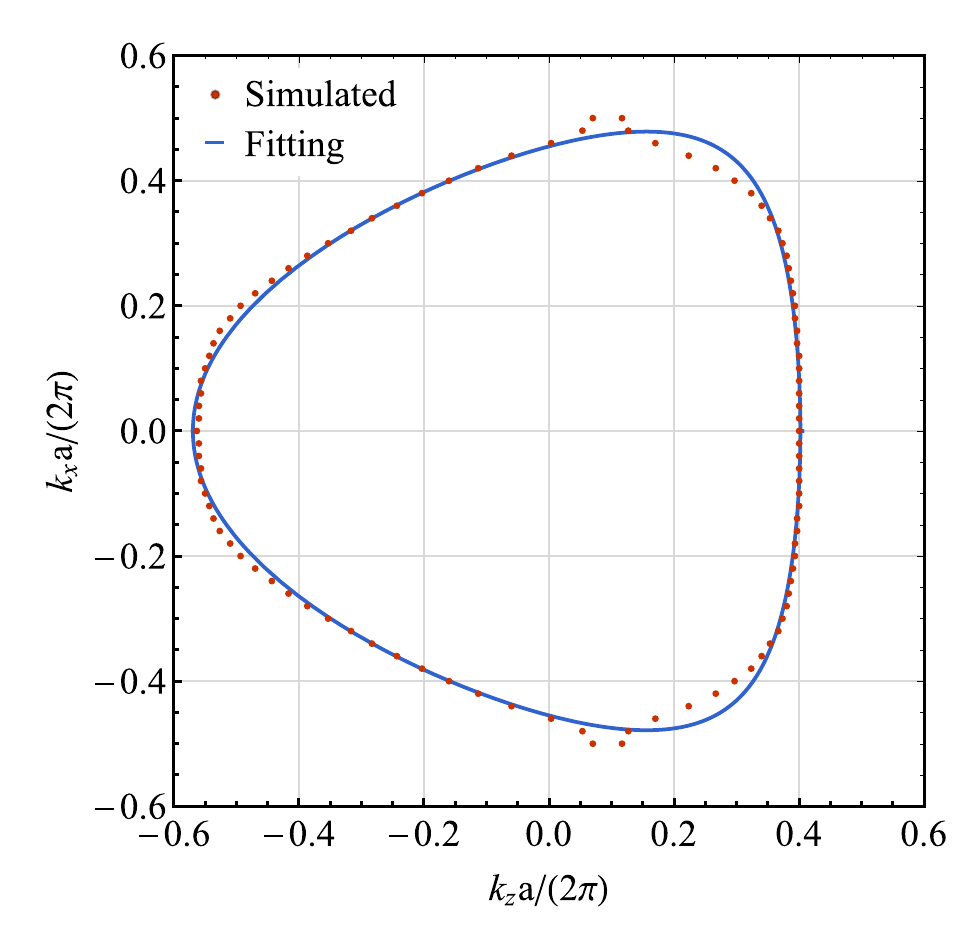}
    \caption{Isofrequency contour for the practical quasi-moving metamaterial described in the End Matter section. Simulated and analytical results with fitted $\chi_\text{qm}$ closely match.}.
    \label{fig:sim isofreq}
\end{figure}

}

\section{Focal length of a lens in a quasi-moving medium}

In this section, we calculate the parameters of a thin lens-shaped cavity cut in the quasi-moving material. We first evaluate the refractive index using the Equation \eqref{eq:wavevec qm}, which exhibits the expected nonreciprocity linear in \(\chi_\text{qm}\):
\begin{equation}
    n(k_x) = \frac{\sqrt{k_z^2 + k_x^2}}{q}=\pm\frac{k_z}{q}\pm\frac{k_x^2}{2k_z\,q}+O(k_x^4) =\sqrt{\eps}+\frac{k_x^2}{q^2\,\eps} \left(\mp \chi_\text{qm}-\frac{\chi_\text{qm} ^2}{2 \sqrt{\eps}}\right)+O\left(k_x^4\right)\:,
\end{equation}
where upper and lower signs correspond to forward ($k_z>0$) and backward ($k_z<0$) propagation, respectively. The term proportional to $\chi_\text{qm}$ is responsible for nonreciprocal dispersion. We recast the above expression in the following form 
\begin{equation}
    n(\theta) = n_\parallel + \zeta \frac{k_x^2}{k^2} =  n_\parallel + \zeta \sin^2 \theta\:, 
\end{equation}
where $\theta$ is the angle between the wave vector and $Oz$ axis, $k^2=k_x^2+k_z^2$,  $n_{\parallel}=\sqrt{\eps}$ and
\begin{equation}
\zeta=\mp\chi_\text{qm}-\frac{\chi_\text{qm}^2}{2\sqrt{\eps}}\:.
\end{equation}
This expression is valid in the paraxial limit in the geometric optics approximation. We assume that \(z\) is the optical axis, an object is located at \((x_\text{o}, 0, z_\text{o})\), the light ray emitted by the object intersects the lens surface at the point \((x_\text{l}, y_\text{l}, z_\text{l}(x_\text{l}, y_\text{l}))\). The surface is spherical, given by the equation
\[
    z_\text{l}(x_\text{l}, y_\text{l}) = R_1 - \sqrt{R_1^2 - x_\text{l}^2 - y_\text{l}^2},
\]
where \(R_1\) denotes curvature radius. Series expansion for optical path length up to the terms quadratic in \(x_\text{o}\), \(x_\text{l}\), \(y_\text{l}\) reads
\begin{equation} \label{eq:optical path}
    \Delta = - n_\parallel z_\text{o} + \frac{n_\parallel + 2 \zeta} {z_\text{o}} x_\text{o} (x_\text{l} - x_\text{o}/2) + \frac{n_\parallel (z_\text{o}/R_1 - 1) - 2 \zeta}{2 z_\text{o}} (x_\text{l}^2 + y_\text{l}^2)\:.
\end{equation}
The surface may separate a medium with parameters \(n_{1\parallel}\), \(\zeta_1\) from another medium characterized by \(n_{2\parallel}\), \(\zeta_2\). In this case we find the location of image by equating coefficients in \eqref{eq:optical path}, namely
\begin{equation} \label{eq:intermediate image}
\eqsys{
    \frac{n_{1\parallel}+ 2 \zeta_1} {z_\text{o}} x_\text{o} &= \frac{n_{2\parallel} + 2 \zeta_2} {z_\text{i0}} x_\text{i0}, \\
    \frac{n_{1\parallel}+ 2 \zeta_1} {z_\text{o}} - \frac{n_{1\parallel}}{R_1} &= \frac{n_{2\parallel}+ 2 \zeta_2} {z_\text{i0}} - \frac{n_{2\parallel}}{R_1},
}
\end{equation}
where \(x_\text{i0}\) and \(z_\text{i0}\) are coordinates of the intermediate image. The lens boundary consists of the two surfaces, on the second surface the image of the intermediate image is cast. We consider a thin lens with \(n_{1\parallel}\), \(\zeta_1\) embedded in the medium possessing \(n_{2\parallel}\), \(\zeta_2\). Applying Eq.~\eqref{eq:intermediate image} twice we obtain
\begin{equation}
\eqsys{
    \frac{n_{1\parallel}+ 2 \zeta_1} {z_\text{o}} x_\text{o} &= \frac{n_{2\parallel} + 2 \zeta_2} {z_\text{i0}} x_\text{i0} = \frac{n_{1\parallel} + 2 \zeta_1} {z_\text{i}} x_\text{i}, \\
    \frac{n_{1\parallel}+ 2 \zeta_1} {z_\text{o}} - \frac{n_{1\parallel}}{R_1} &= \frac{n_{2\parallel}+ 2 \zeta_2} {z_\text{i0}} - \frac{n_{2\parallel}}{R_1}, \\
    \frac{n_{2\parallel}+ 2 \zeta_2} {z_\text{i0}} - \frac{n_{2\parallel}}{R_2} &= \frac{n_{1\parallel}+ 2 \zeta_1} {z_\text{i}} - \frac{n_{1\parallel}}{R_2},
}
\end{equation}
where \(x_\text{i}\) and \(z_\text{i}\) indicate the position of the final image, \(R_2\) denotes the radius of the rightmost surface. From these equations we extract the thin lens formula as well as image magnification. The thin lens formula is similar to the case of a lens made with isotropic materials, but with a modified focal distance:
\begin{equation} \label{eq:focal distance result}
    \frac{1}{z_\text{i}} - \frac{1}{z_\text{o}} = \frac{n_{2\parallel} - n_{1\parallel}}{n_{1\parallel} + 2 \zeta} \left(\frac{1}{R_1} - \frac{1}{R_2} \right)
\end{equation}
Magnification of a lens coincides exactly with the classical result. It equals the ratio between distances from the center of the lens to an object and its image:
\begin{equation}
    \frac{x_\text{i}}{x_\text{o}} = \frac{z_\text{i}}{z_\text{o}}
\end{equation}
The equation \eqref{eq:focal distance result} is used in the main text to showcase the effect of non-reciprocal lensing.

\section{Dipole radiation in a quasi-moving medium}
In this section we investigate the dipole radiation in a quasi-moving medium focusing on the asymmetry of the radiation pattern. We consider a point electric dipole ${\bf d}$
oscillating with the frequency \(\omega\), corresponding to charge density and current 
\[
    \rho = d \, \delta(x) \, \delta(y) \, \delta'(z), \quad \v{j} = i \omega \v{d} \, \delta(x) \, \delta(y) \, \delta(z).
\]
We apply Fourier transformation to Maxwell's equations, obtaining
\begin{align*}
    i \v{k} \times \v{B} + i q \v{D} &= \frac{4 \pi}{c} \frac{i \omega \v{d}}{(2\pi)^3}, \\
    i \v{k} \times \v{E} - i q \v{B} &= 0.
\end{align*}
The amplitude of electric field is found by solving the linear system
\[
    \op{M} \v{E}(\v{k}) = \frac{1}{2 \pi^2} \v{d}, \quad \op{M} = \frac{{\v{k}^\times}^2}{q^2} + \op{\eps} =\frac{\v{k} \otimes \v{k} - k^2 \op{1}}{q^2} + \op{\eps}.
\]
Inverse Fourier transformation yields
\[
    \v{E}(\v{r}) = \frac{1}{2 \pi^2} \int d^3 \v{k}\, e^{i \v{k} \v{r}} \op{M}^{-1}  \v{d}.
\]
The matrix \(\op{M}\) is singular at the isofrequency surface. We adjust the poles assuming the oscillations start  adiabatically in the past and replace \(q\) with \(q + i0\). If \(\v{k} = \v{k}_\parallel + \v{k}_\perp\), \(\v{k}_\parallel \parallel \v{r}\) and \(\v{k}_\perp \perp \v{r}\), then
\[
    \v{E}(\v{r}) = \int d^2 \v{k}_\perp \int d k_\parallel \, e^{i k_\parallel r} \frac{1}{\det \op{M}} \operatorname{adj}(\op{M}) \frac{1}{2 \pi^2} \v{d} = \int d^2 \v{k}_\perp e^{i k_{\parallel i} r} \, 2 \pi i \left(\res_{k_\parallel = k_{\parallel i}} \frac{1} {\det \op{M}} \right) \operatorname{adj}(\op{M}(k_{\parallel i})) \, \frac{1}{2 \pi^2} \v{d},
\]
where \(k_{\parallel i}(k_\perp)\) is the wave-vector projection when the vector lies on the isofrequency surface, \(\operatorname{adj}(\op{M})\) denotes  the adjugate matrix to \(\op{M}\). This evaluation is only applicable when the rank of \(\op{M}\) is 2 on the isofrequency surface, which is the case when \(\chi_{qm} \neq 0\) and \(\v{k}\) is not parallel to \(z\) axis. Higher degeneracies render this particular method inapplicable, but path integrals of the matrix elements of \(\op{M}^{-1}\) may still be evaluated. Further analytical integration of the given expression is not viable in the general case, however far field calculations can be done using stationary phase approximation. Whenever \(r\) is large, the rapid oscillations of \(e^{i k_{\parallel i} r}\) average out to zero, except for the vicinities of local extrema of \(k_{\parallel i}\).  For these points,  the series expansion
\[
    k_{\parallel i} = k_{\parallel i 0} + \frac{1}{2} \left(\frac{\partial^2 k_{\parallel i}}{\partial k_{\perp 1}^2} k_{\perp 1}^2 + \frac{\partial^2 k_{\parallel i}}{\partial k_{\perp 2}^2} k_{\perp 2}^2\right) + O(|\v{k}_{\perp}|^3)
\]
is used, where \(k_{\perp 1}\) and \(k_{\perp 2}\) are the coordinates of \(\v{k}_{\perp}\) in an orthonormal basis diagonalizing the quadratic approximation of the dependency \(k_{\parallel i}(\v{k}_{\perp}) \) . Exponential functions of quadratic expressions are readily integrated using an analytical result
\[
    \int_{-\infty}^\infty e^{i b x^2 / 2} dx = \sqrt{\frac{2 \pi i}{b}}.
\]
Thus, we recover the final expression for the far-field amplitude:
\[
    \v{E}(\v{r}) = \left. -\frac{2 e^{i k_{\parallel} r}}{r}
    \left(\frac{\partial \det \op{M}}{\partial k_{\parallel}}\right)^{-1} \left( \frac{\partial^2 k_{\parallel i}}{\partial k_{\perp 1}^2}  \frac{\partial^2 k_{\parallel i}}{\partial k_{\perp 2}^2} \right)^{-1/2} \operatorname{adj}(\op{M}) \, \v{d} \right|_{k_\perp=0, k_{\parallel} = k_{\parallel i 0}} .
\]
The second derivatives combine to the Gaussian curvature of the isofrequency surface, which we denote by \(K\).
\[
    K = \frac{\partial^2 k_{\parallel i}}{\partial k_{\perp 1}^2}  \frac{\partial^2 k_{\parallel i}}{\partial k_{\perp 2}^2}.
\]
Both electric and magnetic fields can be computed, considering the Fourier image of \(\v{H}\) reads \(\v{H}(\v{k}) = - ( \v{k}^\times / {q} ) \v{E}(\v{k})\).
\begin{align}
    \v{E}(\v{r}) = \left. - \frac{2 e^{i k_{\parallel} r}}{\sqrt{K} r}
    \left(\frac{\partial \det \op{M}}{\partial k_{\parallel}}\right)^{-1}  \operatorname{adj}(\op{M}) \, \v{d} \right|_{k_\perp=0, k_{\parallel} = k_{\parallel i 0}}, \\
    \v{H}(\v{r}) = \left. \frac{2 e^{i k_{\parallel} r}}{\sqrt{K} r}
    \left(\frac{\partial \det \op{M}}{\partial k_{\parallel}}\right)^{-1}  \frac{\v{k}^\times}{q}\operatorname{adj}(\op{M}) \, \v{d} \right|_{k_\perp=0, k_{\parallel} = k_{\parallel i 0}}.
\end{align}

These results are enough to find the Poynting vector, which acquires a spatial dispersion-induced correction in our case \cite{Landau}
\[
    \v{S} = \frac{c}{8 \pi} \left( \Re \v{E}^* \times \v{H} - \frac{q}{2} \frac{\partial \eps_{ij}}{\partial \v{k}} E_i^* E_j \right)
\]
The angular distribution of radiation power can be of interest. We evaluate it as a Poynting vector flux through a section of a distant sphere
\[
    \frac{\delta P}{\delta \Omega_S} = \frac{r^2 \delta \Omega_r \v{S} \cdot \v{n}_r }{\delta \Omega_S},
\]
where \(\Omega_r\) and \(\Omega_S\) are solid angles for position on the sphere and Poynting vector respectively, \(\v{n}_r = \v{r}/r\). We write
\begin{align*}
    \delta \Omega_S = \sin \theta_S \, \delta \theta_S \, \delta \phi_S, \\
    \delta \Omega_r = \sin \theta_r \, \delta \theta_r \, \delta \phi_r,
\end{align*}
arriving to the expression for the power distribution in terms of the Poynting vector and the Jacobian of the solid angle mapping
\[
    \frac{\delta P}{\delta \Omega_S} = r^2  \v{S} \cdot \v{n}_r \frac{\partial (\cos \theta_r, \phi_r) }{\partial (\cos \theta_S, \phi_S) }.
\]
\begin{figure}
    \centering
    \includegraphics[width=0.9\linewidth]{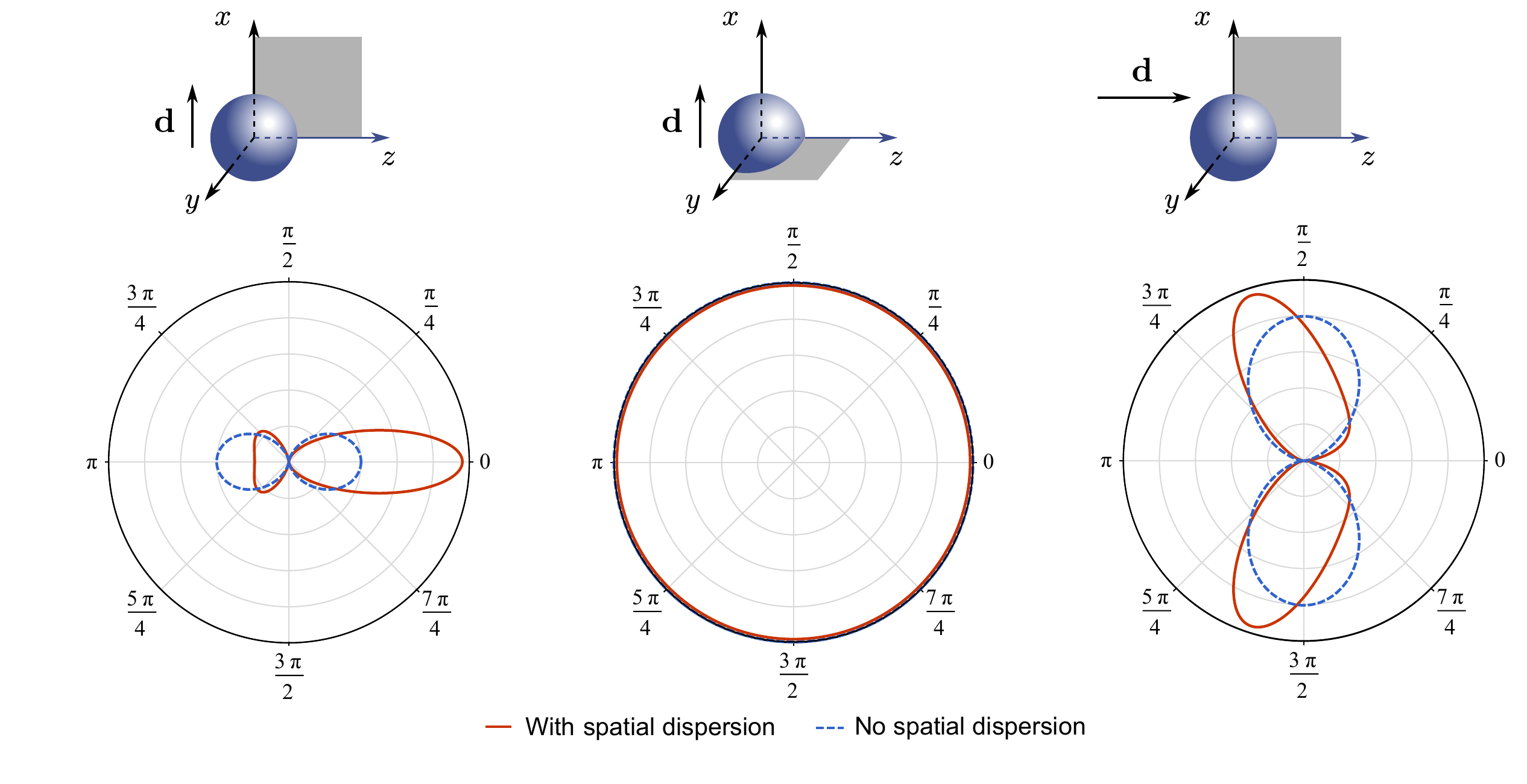}
    \caption{Radiation patterns of a point dipole in the far field, depending on the dipole orientation, with the magnitude of non-reciprocity \(\chi_\text{qm} = -0.2\).}.
    \label{fig:rad pattern}
\end{figure}
Evaluated distributions are presented in Figure~\ref{fig:rad pattern}. If dipole is oscillating along \(z\), the rotational axis of the material, we observe backscattering suppression due to the nonreciprocity [Fig.~\ref{fig:rad pattern}, right]. The pattern is rotationally symmetric. 

The dipole oriented orthogonally to the rotation axis has a more complex asymmetric radiation pattern [Fig.~\ref{fig:rad pattern}, left]. Two cross-sections are shown in the figure. In the plane \((xz)\) the nonreciprocity is apparent, however the quasi-moving nonlocality does not affect the radiation power distribution in the \((yz)\) plane. In turn, asymmetric radiation pattern of a dipole ensures that the backscattering can be reduced in the entire bulk of the medium.

Notably, the radiation diagram for the dipole perpendicular to the symmetry axis is discontinuous at the limit \(\v{k} \to (0, 0, q)^\transp\) -- hence the radiation power has different values near the axis \(z\) for the first two panels of the figure. This is caused by eigenmode degeneracy when \(\v{k} =(0, 0, q)^\transp \). In the vicinity of this vector the degeneracy breaks in different ways, depending on the direction of the wave vector. For instance, if \(k_x = 0\) and \(k_y \neq 0\), the ordinary mode is polarized along the axis \(y\), and the extraordinary mode is polarized along \(x\); the dipole excites the extraordinary mode, affected by spatial dispersion. The situation is opposite when  \(k_y = 0\) and \(k_x\neq 0\), in this case the emission is accomplished via the unchanged ordinary mode. Because these radiation channels have very different efficiency, owing to the difference in the curvatures of the isofrequency surfaces, the power noticeably varies when the observation point rotates around~\(z\).

The effect of discontinuity in the radiation pattern is not unique to the description of quasi-moving dispersion, a similar discontinuity is predicted for uniaxial crystals~\cite{Hayat2019} when the dipole is not parallel to the anisotropy axis. It is instructive to outline similarities between uniaxial crystals and the quasi-moving media in the formalism of scalar Green's functions, used in the paper. For that, we recast the matrix from Maxwell's equations  in the form

\[
    \op{M}^{-1} = \frac{q^2}{d_1(\v{k})} \op{1} + \frac{1}{d_1(\v{k}) \, d_2(\v{k})} \left( n_1(\v{k})\, \v{e}_z \otimes \v{e}_z  + n_2(\v{k})\, \v{k} \otimes \v{k} + n_3(\v{k})\, [\v{k} \otimes \v{e}_z + \v{e}_z\otimes \v{k} ] \right),
\]
where
\begin{align*}
    d_1(\v{k}) &= q^2 - \v{k}^2, \\
    d_2(\v{k}) &= q^3 - q k_z^3 - (k_x^2 + k_y^2)(q + q \chi_\text{qm}^2 + 2 \chi_\text{qm} k_z), \\
    n_1(\v{k}) &= q^3 \chi_\text{qm} (\chi_\text{qm} (k_x^2 + k_y^2) + k_z (2 q + \chi_\text{qm} k_z) ), \\
    n_2(\v{k}) &= q (q^2[\chi_\text{qm}^2 - 1] + k_x^2 + k_y^2 + 2 q \chi k_z +k_z^2), \\
    n_3(\v{k}) &= q^2 \chi_\text{qm}  \left(k_z q \chi_\text{qm} - k_x^2 - k_y^2 + k_z^2 + q^2\right).
\end{align*}
The Fourier image of \(\op{M}^{-1}\) is the dyadic Green's function, expressed as
\[
    \op{G}^{\text{ee}}(\v{r}) = q^2 g_1(\v{r}) \, \op{1} + \left\{ \v{e}_z \otimes \v{e}_z \, n_1(-i \nabla)  -  \nabla \otimes \nabla \, n_2(-i\nabla) - i [\nabla \otimes \v{e}_z + \v{e}_z \otimes \nabla]  \, n_3(-i \nabla) \right\} g_3(\v{r})
\]
via scalar Green's functions 
\[
    g_1(\v{r}) = \int d^3 \v{k} \frac{1}{d_1(\v{k})} e^{i \v{k} \v{r}}, \quad g_2(\v{r}) = \int d^3 \v{k} \frac{1}{d_2(\v{k})} e^{i \v{k} \v{r}}, \quad g_3(\v{r}) = \int d^3 \v{k} \frac{1}{d_1(\v{k}) d_2(\v{k})} e^{i \v{k} \v{r}} = (g_1 * g_2)(\v{r})
\]
Here the operation \(*\) denotes convolution. The mixed derivatives \(\partial_x \partial_y\) of the function \(g_3\) are  not continuous at \(r_x=r_y=0\) (this parts acquires contributions from the wave vectors near \( (0,0,q)^\transp \)), and as a consequence the radiated field amplitude is discontinuous as well. This method would provide an analytical near-field solution, if the expression for \(g_3\) was analytically integrable.


\section{Details on backscattering suppression in a realistic microwave metamaterial}

{
In this section, we provide additional data relevant to the analysis of backscattering suppression in a realistic microwave metamaterial discussed in the End Matter section. In addition to calculating the change of forward transmittance, we provide quantitative analysis of reflectance. 

In contrast to the transmittance properties [Fig.~\ref{fig:sim refl} (a)], reflectance of an idealized structure is the same in forward and backward directions [Fig.~\ref{fig:sim refl} (b)]. This property is guaranteed by the symmetry of a unit cell with respect to reflection in \(Oxz\) plane accompanied by the time reversal. Attenuation of the waves due to complex $k_z$ affects the reflectance plots by decreasing the amplitude of Fabry-Perot oscillations.  Introducing defects within the sample affects both transmittance and reflectance [Fig. \ref{fig:sim refl} (c,d)], the change is most pronounced near the extrema of oscillations. When the frequency approaches 1.15 GHz, the influence on both reflectance and transmittance diminishes. 

We therefore conclude that the proposed photonic structure not only preserves transmission regardless of the defects, but also acquires robust reflection. The waves scattered backwards are absorbed, and reflectance properties are affected only by the boundary conditions. To quantify this property, we use the same methodology as in the End Matter, and analyse mean squared deviation \(\text{MSD}(R)\) [Fig. \ref{fig:sim refl} (f)]. Without significant changes to the absolute values of reflectance, the influence of defects drops from 25\% to just 5\%. Thus, backscattering suppression in the quasi-moving metamaterial is manifested both in reflection and transmission.

\begin{figure}[hb]
    \centering
    \includegraphics[width=0.9\linewidth]{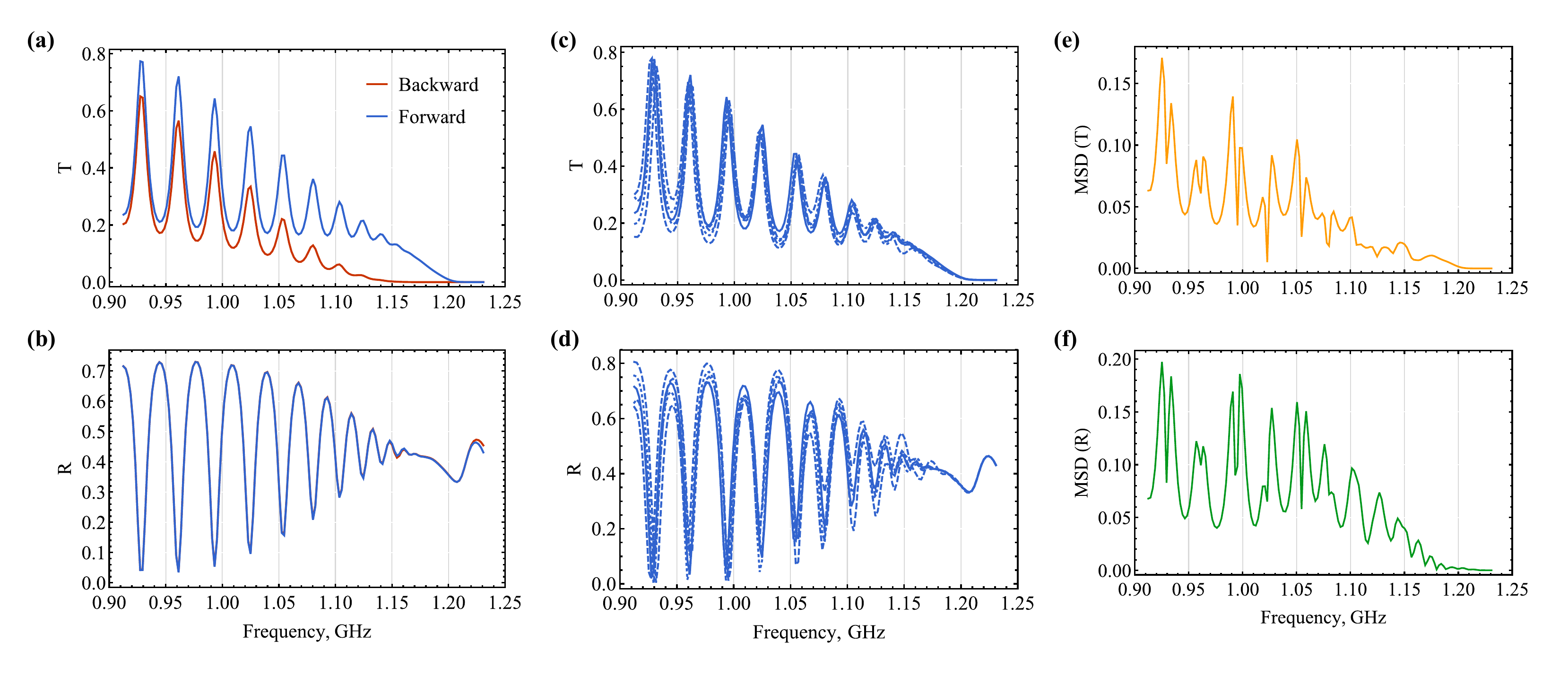}
    \caption{(a,b) Scattering parameters of a finite sample consisting of 16 unit cells. (c,d) Forward transmittance and reflectance for a perfect crystal  (solid line) and modifications caused by the different types of defects (dashed and dotted lines). (e,f) Mean squared deviations of the scattering parameters caused by the defects.}
    \label{fig:sim refl}
\end{figure}

}

\bibliography{NonReciprocalLens}